\documentclass[preprint]{elsarticle}

\usepackage[headsep=.2in,footskip=.2in]{geometry}
\usepackage[bookmarks=false]{hyperref} 
\hypersetup{
	colorlinks=true,
	linkcolor=blue,
	filecolor=magenta,      
	urlcolor=cyan,
	citecolor=blue,
}
\usepackage{booktabs}
\usepackage{titlesec} 

\usepackage{paracol} 
\columnratio{0.6,0.4} 
\titlespacing\section{0pt}{6pt plus 4pt minus 2pt}{4pt plus 2pt minus 2pt}
\titlespacing\subsection{0pt}{6pt plus 4pt minus 2pt}{4pt plus 2pt minus 2pt}
\titlespacing\subsubsection{0pt}{6pt plus 4pt minus 2pt}{4pt plus 2pt minus 2pt} 

\titleformat{\section}
{\large\bfseries}
{\thesection.}{0.5em}{} 

\titleformat{\subsection}
{\normalsize\bfseries\itshape}
{\thesubsection.}{0.5em}{} 

\titleformat{\subsubsection}
{\normalsize\itshape}
{\thesubsubsection.}{0.5em}{} 

\usepackage{fancyhdr}
\usepackage{color}

\usepackage{graphicx} 
\graphicspath{ {images/} }
\usepackage[font=normalsize,skip=0pt]{caption}
\usepackage{subcaption}
\usepackage{wrapfig} 
\usepackage{multirow}
\usepackage{array} 

\usepackage{amsmath}
\usepackage{enumitem} 
\usepackage{amssymb} 
\usepackage[mathcal]{euscript}

\usepackage[british]{babel}
\usepackage{natbib} 

\usepackage{ifthen} 
\usepackage{lscape} 

\usepackage{pifont} 
\usepackage[usenames,svgnames,table]{xcolor} 
\definecolor{shade}{HTML}{F5DD9D} 

\let\OLDthebibliography\thebibliography
\renewcommand\thebibliography[1]{
	\OLDthebibliography{#1}
	\setlength{\parskip}{0pt}
	\setlength{\itemsep}{0pt plus 0.3ex}
}

\usepackage{fancyhdr}
\usepackage{algorithm,algpseudocode}

\newcolumntype{C}[1]{>{\centering\arraybackslash}p{#1}}
\newcolumntype{L}[1]{>{\arraybackslash}p{#1}}

\definecolor{LightRed}{rgb}{1,0.88,1}
\definecolor{LightGreen}{rgb}{0.88,1,0.88}
\definecolor{Cyan}{rgb}{0.88,1,1}
\definecolor{Black}{rgb}{0.0, 0.0, 0.0}
\definecolor{Blue}{rgb}{0.2, 0.6, 1}
\definecolor{Yellow}{rgb}{1, 1, 0}
\definecolor{DarkYellow}{rgb}{0.922, 0.773, 0.176}
\definecolor{Red}{rgb}{1, 0, 0}
\definecolor{Orange}{rgb}{1, 0.6, 0.2}
\definecolor{Green}{rgb}{0.2,0.8,0.2}
\definecolor{Purple}{rgb}{0.6, 0.4, 1}
\definecolor{Grey}{RGB}{179, 179, 179}

\newsavebox{\verbbox} 

\usepackage{tikz} 
\usepackage{calc} 
\usetikzlibrary{calc,arrows}

\usepackage{amsmath}

\usepackage{amssymb}

\biboptions{authoryear}

\usepackage[figuresright]{rotating}
\newtheorem{definition}{Definition}[section]

\usepackage[bookmarks=false]{hyperref}
    \hypersetup{colorlinks,
      linkcolor=blue,
      citecolor=blue,
      urlcolor=blue}

\usepackage[utf8]{inputenc}

\makeatletter
\def\ps@pprintTitle{%
	\let\@oddhead\@empty
	\let\@evenhead\@empty
	\let\@oddfoot\@empty
	\let\@evenfoot\@oddfoot}
\makeatother

\providecommand{\toprule}{\hline}

\providecommand{\cmidrule}[1]{\hline}
\providecommand{\colrule}{\hline}
\providecommand{\botrule}{\hline}

\begin{document}
\begin{frontmatter}



\title{Prediction of airport on-time performance}


\author[a]{Xavier Lemay} 
\author[a]{Fabian Bastin\corref{cor1}}

\address[a]{University of Montreal, Department of Computer Science And Operations Research, CP 6128 Succ Centre-Ville, Montreal, QC, Canada  H3C 3J7}

\begin{abstract}
We investigate the factors contributing to departure and arrival delays at a major international airport and develop predictive models to estimate both the likelihood and duration of delays. Using logistic regression, random forest, and gradient boosting methods, we identify key predictors of flight punctuality, including historical delay rates of flight numbers and airlines, weather conditions, runway traffic, walk time from security to gate, and overall airport congestion. Our models achieve strong inference and predictive performance in both classification and regression tasks, demonstrating the potential for targeted operational interventions to improve on-time performance and providing actionable insights for airport management and airline operations.
\end{abstract}

\begin{keyword}
On-time performance; Flight delay prediction; Forecasting; Airports; Machine learning; Regression analysis; Classification models; Data inference; Explainability




\end{keyword}
\cortext[cor1]{Corresponding author. Email: bastin@iro.umontreal.ca}
\end{frontmatter}


\section{Introduction}
\label{introduction}

Flight delays are a common issue in the aviation industry, affecting both airlines and passengers. Delays can lead to increased operational costs for airlines~\citep{Balletal10}, missed connections for passengers, and overall dissatisfaction with air travel~\citep{KimPark16}.
A key metric to measure the performance of airlines and airports is the On-Time Performance (OTP), defined as the percentage of flights that depart or arrive within 15 minutes of their scheduled time. Similarly, a flight is said to respect its OTP criterion if it departs or arrives within 15 minutes of the scheduled time.
OTP is a critical factor in customer satisfaction and choice, with customers being willing to pay a premium to reduce the likelihood of delays~\citep{Yimg17}.

However, there is a significant gap in understanding factors that lead to delays, especially from the point of view of airports~\citep{WangBiXieZhao22} and the interaction with airlines~\citep{CalzFage23}.
This paper aims to contribute to bridging that gap by providing an analysis of factors leading to delays at arrivals and departures from a major international airport, and by developing predictive models to forecast delays.
We utilize a comprehensive dataset that includes flight schedules, weather conditions, and operational data from the airport. By applying tree-based techniques, we identify key predictors of delays and develop models that can accurately forecast delays.
We also use forecasted weather data to simulate real-world conditions under which these models would operate.

The rest of the paper is organized as follows. Section~\ref{sec:literature} presents a brief literature review, followed by a description of the methodology in Section~\ref{sec:methodology}. The data used in this study are described in Section~\ref{sec:data}. The results are discussed in Section~\ref{sec:results}, and applications to airport operations are presented in Section~\ref{sec:application}. Finally, Section~\ref{sec:conclusion} concludes the paper and outlines directions for future research.


\section{Literature Review}
\label{sec:literature}

Predicting flight on-time performance (OTP) and delays has been widely studied in the fields of air transportation, operations research, and applied machine learning. Early research relied on statistical and econometric models to analyze the impact of weather, scheduling practices, and traffic congestion on flight delays~\citep{BubaGagg21, Carvetal21}. While these approaches offered interpretability, they struggled with nonlinear and high-dimensional data. 

With the growth of large-scale operational and weather datasets, machine learning methods have become prominent~\citep{WandChenSun25}. Tree-based ensemble methods such as Random Forests and Gradient Boosting Machines have consistently demonstrated strong performance and the ability to identify key predictors, including weather variables, traffic congestion, and propagation of prior delays~\citep{Etan19, Dai24, LiJing22}. Hybrid and optimized ensemble models further enhance predictive accuracy, particularly in handling class imbalance in delay datasets~\citep{Dai24}. 

More recent work addresses the networked nature of delay propagation. Spatio-temporal random forests~\citep{GuoYuHaoWangJianZong21, LiJing22} and graph-based neural networks~\citep{Zhenetal24, KhanMaChunWen21} explicitly model airport connectivity and flight dependencies, capturing cascading delay effects across the network. Probabilistic approaches have also emerged to improve calibration and provide actionable forecasts for operational decision-making~\cite{BeltRibedeWiSun25}.

A systematic review and a taxonomy of these various approaches is proposed by \citet{Carvetal21}.
Across these studies, weather, prior leg delays, and traffic demand consistently emerge as critical predictors. Ensemble methods provide robust baselines with interpretable feature importances, while deep and graph-based approaches show promise for longer-horizon, network-level forecasting. Nevertheless, challenges remain in ensuring model interpretability, standardizing evaluation metrics, and translating predictions into effective operational interventions.

While airline delays have received more attention, a few studies specifically focus on airport delays~\citep{ChinMeilMurpThya13}.
\cite{XuSherLask08} develop piecewise linear regression models to estimate generated and absorbed delays, and found that the major causes inbound delay, scheduled turnaround time, and carrier delay.
\cite{ChanTran18} analyze the resilience of an air transportation network using publicly available data while  \cite{TanLuWang25} focus on the delay propagation and the real-time detection of delays.
A comprehensive review of delay propagation among airports is provided by \cite{LiYao25}.

A better understanding of the factors affecting the OTP offer opportunities to improve it and reduce delays.
On one side, airline can reduce flight delays and improve their arrival OTP by padding their schedules~\citep{YimgGorj19}, but also with active strategies that will lead to higher customer satisfaction~\citep{HajkBada20, MancSohoDesh23}.
On the other side, improving the departure OTP often requires a review of the airport operations, and airport management can mitigate or worsen the propagation of delays~\citep{RodrGomeArnaPereBarr19, WangAlliBarnGondGottManc22, YangChenHuSongMao23}.
A first step is therefore to identify the causes of delays and OTP degradation, and to isolate the control variates since, as observed by \cite{Carvetal21} flight delay estimation can support tactical and operational decisions of airline managers and airport operators~\citep{RibeTayNgBiro25}.

Especially, a better understanding and prediction of arrival and departure delays on various temporal scales (short-term and long-term) at a given airport can help the manager to identify and anticipate operating conditions creating or propagating delays.
Using data provided by a major international airport, and leveraging on the recent success of machine learning techniques, this study aim to provide a more comprehensive analysis of factors affecting a specific airport along as more accurate forecasts, involving individual flight delays on the short term and average OTP on the long-term.

\section{Methodology}
\label{sec:methodology}

Our main objective is to forecast the delay (arrival or departure) up to a given horizon $T$ in advance, and to understand the main drivers affecting the delays, allowing real-time and preemptive actions.
The on-time performance (OTP) can also be derived by setting a threshold on the delay under which we can consider the flight to be on time.
The delay is formally computed as the time elapsed between the scheduled time and the moment when, for a departure, the plane leaves the gate, even if it not by its own means (for instance using a pushback tractor), or arrives at its assigned gate for an arrival.
More formally, we introduce the following definitions.

\begin{definition}[Times and Delay]
	Let \(F=\{1,\dots,N\}\) be the set of flights considered in a given period.
	For each flight \(i\in F\), denote
	\begin{itemize}
		\item \(s_i\): the \emph{scheduled} time of an event (departure or arrival),
		\item \(a_i\): the \emph{actual} time of that event,
		\item the delay \(d_i := a_i - s_i\) (measured in minutes). By convention, \(d_i\) can be negative (early operation).
	\end{itemize}
	\label{def:delay}
\end{definition}

	\begin{definition}[OTP with Threshold]
	Fix a tolerance threshold \(\tau \ge 0\) (in minutes).
	A flight \(i\) is said to be \emph{on-time} if \(d_i \le \tau\).
	The \emph{On-Time Performance} (OTP) of the set \(F\) at threshold \(\tau\) is defined as
	\begin{equation}
	\operatorname{OTP}_{F}(\tau)
	\;:=\;
	\frac{1}{N}\sum_{i\in F} \mathbf{1}\{d_i \le \tau\},
	\label{eq:otp}
	\end{equation}
	where \(\mathbf{1}\{\cdot\}\) is the indicator function. Thus \(\operatorname{OTP}_F(\tau)\in[0,1]\) is the proportion of flights classified as \emph{on-time} according to threshold \(\tau\).
	\label{def:otp}
\end{definition}

We set $\tau=$ 15 minutes, as it is usually done~\citep{Euro05}, and divide the models in two sets, reflecting their purpose.
We first consider a class of \textit{inference} models,
used to identify key features that affect delays.
For such models, we do not use ``proxy features'', that is, features which do not have a direct effect on flights, for instance
the historical OTP of a flight number, as this feature is correlated with delays, but does not cause delay.
Since inference models are meant to capture the origin of delays, we allow ourselves to use ``instant'' features, that is, features that would not be known at the time of prediction, but that \textit{do} have a direct impact on delays, for instance the actual weather at the time of departure/arrival.
The second set of models is composed of \textit{prediction} models, aiming to forecast the delay as accurately as possible given the information available at the time of prediction.
Therefore, we allow the use of proxy features, but remove features not known at the prediction time.
In particular, we use forecasted weather instead of actual weather, and allow the use of the predicted airport OTP~\eqref{eq:otp} for the period of interest, and the historical OTP of the flight number, airline, and gate.
Thus, we do not use the same set of features for both kind of models. 

We furthermore divide these two sets in classification and regression models.
Classification models predict if a flight will be on-time (as in Definition~\ref{def:otp}) 
while regression models predict the delay duration (in minutes).
We use various metrics to evaluate the performance of the models developed in this paper, depending on their purpose.
For classification tasks, 
we rely on precision, recall, and $F_1$-score, defined as
\begin{align*}
	\text{precision} &= \frac{TP}{TP+FP}, \\
	\text{recall} &= \frac{TP}{TP+FN}, \\
	F_1\text{ score} &= 2\frac{\text{precision}\times\text{recall}}{\text{precision}+\text{recall}},
\end{align*}
where $TP$, $FP$, and $FN$ stand for true positive, false positive, and false negative, respectively. The $F_1$-score is the harmonic mean of precision and recall.


For a binary classification problem with classes $A$ (on-time) and $B$ (delayed), metrics such as precision, recall, or $F_1$-score can be averaged as follows:
\begin{itemize}
	\item \textbf{Macro Average}: Compute the metric for each class and take the simple mean. Each class has equal weight.
	\[
	\text{MacroAvg}(M) = \frac{M_A + M_B}{2}
	\]
	
	\item \textbf{Weighted Average}: Compute the metric for each class and take a mean weighted by the number of instances in each class.
	\[
	\text{WeightedAvg}(M) = \frac{n_A \cdot M_A + n_B \cdot M_B}{n_A + n_B}
	\]
	where $n_A$ and $n_B$ are the numbers of instances in classes $A$ and $B$, respectively.
\end{itemize}
Macro average assigns equal importance to both classes, whereas weighted average accounts for the relative sizes of the classes in the dataset. In particular, the number of delayed flights is typically much smaller than the number of on-time flights. However, accurately identifying delayed flights is of primary importance. Therefore, in our experiments, we favor the use of weighted averages to better reflect class distribution while preserving sensitivity to the minority class.

The quality of regression models is measured by computing the mean average error (MAE) and the root mean square error (RMSE), defined as
\begin{align}
	MAE &= \frac{1}{N} \sum_{i = 1}^N | \hat{y}_i - y_i |, \label{eq:MAE} \\
	RMSE &= \sqrt{ \frac{1}{N} \sum_{i = 1}^N \left( \hat{y}_i - y_i \right)^2 }, \label{eq:RMSE}
\end{align}
where $N$ is the number of observations, $y_i$ is the $i^{th}$ ground truth value (e.g., actual airport delays) and $\hat{y}_i$ is the  $i^{th}$ predicted value (e.g., predicted airport delays).

We now provide more details about the technique used throughout this paper.

\subsection{Random forests}

Random forests are an ensemble learning method for both classification and regression tasks. 
They operate by constructing a large number of decision trees during training and producing a final output by aggregating the predictions of the individual trees: the \emph{mode} of the classes for classification, or the \emph{mean} prediction for regression \citep{Brei01,LiawWien02}.  

Importantly, because the classification output is based on the mode of the predicted classes, it is possible to extract a \emph{predicted probability} for a flight satisfying its on-time performance (OTP) criterion by computing the proportion of trees that predict the flight will meet the OTP threshold. 
This probabilistic output provides more informative insights than a simple binary prediction, enabling the development of tools to assess the \emph{likelihood of a flight being delayed} \citep{ZhonYuHuanZhan25,WangBiXieZhao22}.  

Each tree in the forest is trained on a \emph{bootstrap sample} of the training data; that is, a random subset of the data selected with replacement. 
This sampling procedure reduces overfitting and improves the generalization ability of the model. 
Each decision tree consists of a sequence of binary splits based on the features of the data, with the goal of partitioning the dataset into subsets that are as homogeneous as possible with respect to the target variable. 
Splits are chosen according to criteria such as \emph{Gini impurity} or \emph{information gain} for classification tasks, and \emph{mean squared error} for regression tasks. 
Additionally, the depth of the trees and the number of features considered at each split can be controlled to further mitigate overfitting and enhance model robustness \citep{Brei01}.  

In practice, the performance of a random forest depends on its \emph{hyperparameters}, including the number of trees, the maximum depth of each tree, the minimum number of samples required to split a node, and the number of features considered at each split. 
These hyperparameters are typically optimized through cross-validation to balance predictive accuracy and computational efficiency \citep{LiawWien02}. 
Furthermore, random forests provide measures of \emph{feature importance}, which quantify the relative contribution of each variable to the model’s predictive performance.

\subsection{Gradient boosting}

Gradient boosting is another ensemble learning method that builds models in a sequential manner. 
Unlike random forests, which train multiple trees independently, gradient boosting constructs trees one at a time, with each new tree attempting to correct the errors made by the ensemble of previously trained trees. 
This is accomplished by fitting the new tree to the \emph{residual errors} of the current model, that results from a combined ensemble of all previously trained trees~\citep{Frie01}.  

The key idea behind gradient boosting is to optimize a specified loss function by adding new models that predict the \emph{negative gradient} of the loss function with respect to the current predictions. 
In doing so, the algorithm iteratively focuses on the most difficult cases, thereby refining performance at each stage. 
Gradient boosting can be applied to both classification and regression tasks, and has been shown to achieve \emph{state-of-the-art performance} across a wide variety of domains, including flight delay and OTP prediction \citep{Frie02,ChenGues16}.  

\subsection{Permutation feature importance}

Permutation feature importance is a widely used technique for assessing the contribution of individual features in a machine learning model. 
The basic idea is to evaluate the change in model performance when the values of a specific feature are randomly permuted, thereby breaking the association between that feature and the target variable. 
If permuting a feature results in a substantial decrease in model performance, it suggests that the feature is important for generating accurate predictions \citep{Brei01,FishRudiDomi19}.  

A key advantage of permutation importance is that it is \emph{model-agnostic}, meaning that it can be applied to any machine learning model, including tree-based ensembles such as random forests and gradient boosting. 
It also provides a more reliable measure of feature relevance compared to split-based importance measures, which are known to be biased toward featureswith more levels or higher cardinality \citep{StroBoulZeilHoth07}.  

However, permutation importance does not provide information about the \emph{direction} of the effect of a feature on the target variable, but only about the strength of its association. 
Thus, complementary techniques (e.g., partial dependence plots, SHAP values, or ICE plots) are required to gain insights into how a feature influences predictions \citep{LundLee17,RudiChenChenHuanSemeZhon22}.  

\section{Data}
\label{sec:data}


We collected flight data from a major international airport, covering all departures and arrivals between 2019 and 2025. For each flight, the dataset includes the flight number, scheduled arrival and departure times, actual arrival and departure times, origin and destination airports, airline, aircraft type, assigned gate (including whether or not it is a remote gate), and the walking time from the security checkpoint to that gate. 

In addition, we collected realized weather data from the same period, including ground temperature, wind speed, visibility, and precipitation amount, updated hourly. Forecasted weather data were also obtained from an industrial partner. These forecasts include wind speed and precipitation, with precipitation further classified into rain, snow, freezing rain, and ice pellets. New forecasts are produced every six hours, each providing hourly predictions for up to 48 hours in the future. To account for potential missing data, we restrict the prediction horizon $T$ in this study to 36 hours, consistent with practices in the aviation forecasting literature \citep{BuxiHans13, ReboBala14}.

Additional information on airlines operating in the dataset was collected independently. This includes the continent of origin, airline size (measured by number of aircraft), number of destinations served, and membership in airline alliances. For each airport in the dataset, we also collected the continent of origin and the distance from the main study airport. Finally, we obtained from the airport authority the average monthly load factor of flights by sector (domestic, transborder, international).


Our raw flight dataset initially contained $963{,}111$ observations. We cleaned the data by removing cancelled or inoperational flights, as well as data from the most critical period of the COVID-19 pandemic (2020-04-01 to 2021-07-01), since flight patterns during this period were atypical and not representative of post-pandemic operations. Cargo, private, and military flights were also excluded. After these filters, $849{,}889$ entries remained.

Among the remaining observations, $5732$ were missing gate numbers. For these flights, we imputed the gate using the most common gate associated with the same flight number. After this procedure, 26 flights remained without a gate assignment. Since the flight numbers were otherwise unique, we interpreted these as data entry errors and removed them. In addition, 33 flights were missing their origin or destination airport. Given the small number of such cases, we deleted these entries as well.

Weather and forecasting data were similarly treated. If an entry was missing, we imputed it by taking the mean of the preceding and following observations. This approach only left gaps longer than one hour. For visibility, we used forward filling, as this variable tends to change gradually. For temperature, precipitation, and wind speed, we imputed missing values using data from the same time of year in the previous year. This seasonal imputation approach is motivated by the fact that these variables vary rapidly within a day, while seasonal patterns remain relatively stable.

After cleaning and imputation, we obtained a final dataset of $652{,}781$ entries. Among these, $195{,}255$ flights ($29.91\%$) experienced a delay, while $457{,}526$ flights ($70.09\%$) were on time. The average delay duration was $20.80$ minutes, with a standard deviation of $66.98$ minutes. The dataset is composed of $312{,}277$ arrivals ($47.84\%$) and $340{,}504$ departures ($52.16\%$). In terms of sector, $305{,}001$ flights ($46.72\%$) were domestic, $131{,}761$ ($20.18\%$) were transborder, and $216{,}019$ ($33.09\%$) were international.

\subsection{Qualitative Analysis of Feature Impacts on OTP}

\begin{figure}[htbp]
	\centering
	\begin{minipage}[b]{0.31\textwidth}
		\centering
		\includegraphics[width=\textwidth]{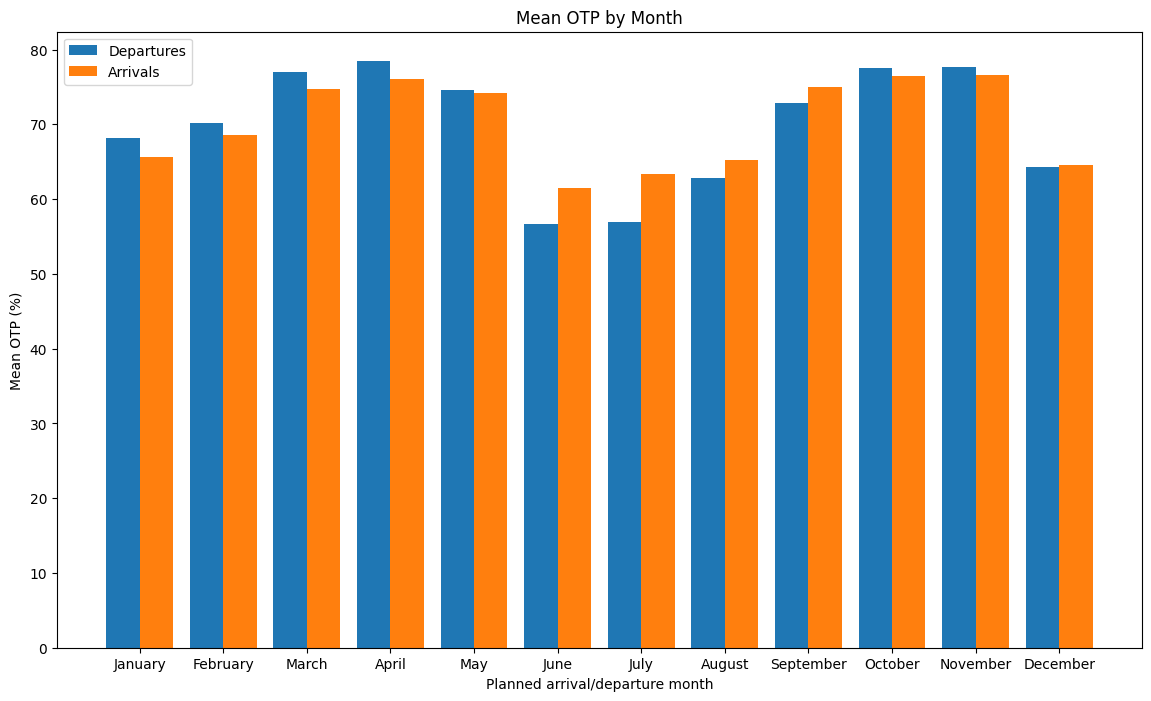}
		\caption{(a) OTP by month of year}
		\label{fig:qualitative_analysis_otp_month}
	\end{minipage}
	\hfill
	\begin{minipage}[b]{0.31\textwidth}
		\centering
		\includegraphics[width=\textwidth]{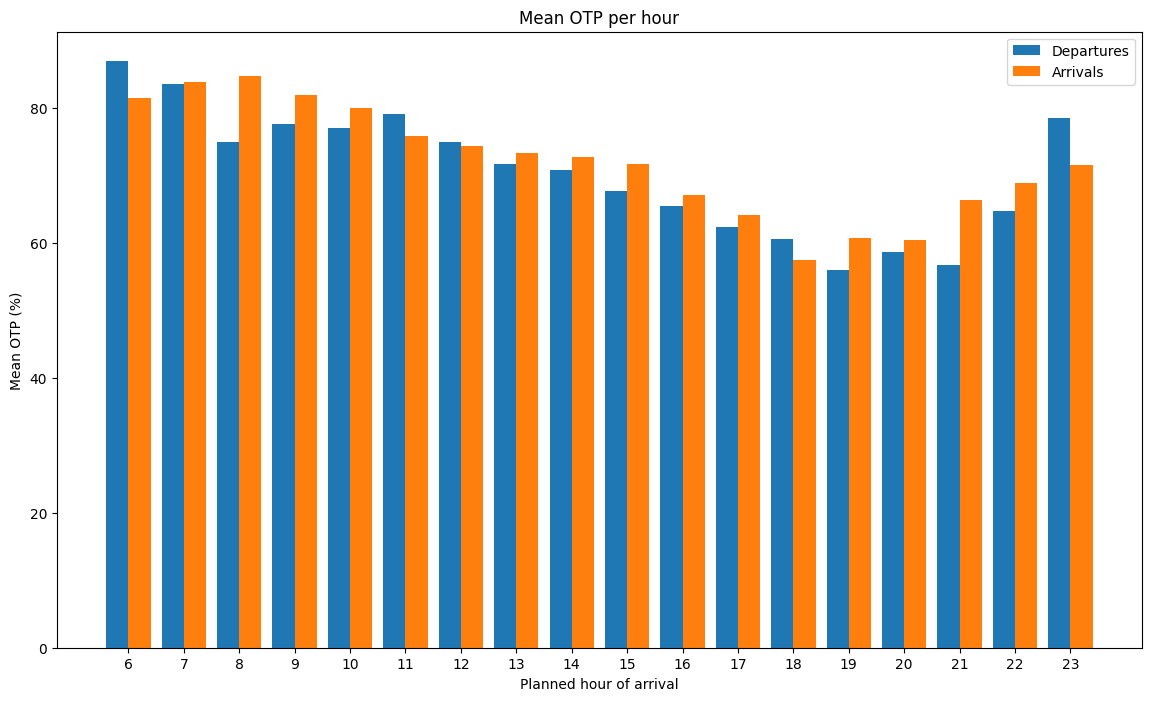}
		\caption{(b) OTP by hour of day}
		\label{fig:qualitative_analysis_otp_hour}
	\end{minipage}
	\hfill
	\begin{minipage}[b]{0.31\textwidth}
		\centering
		\includegraphics[width=\textwidth]{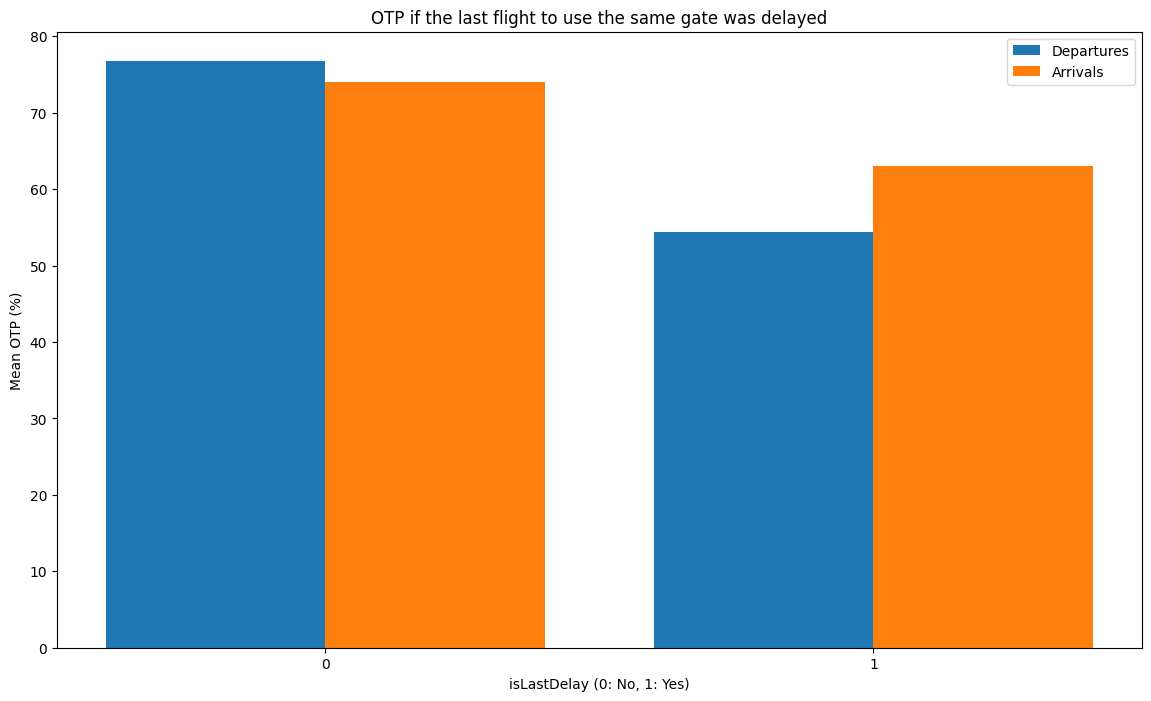}
		\caption{(c) OTP if the last flight using the same gate was delayed or not}
		\label{fig:qualitative_analysis_otp_last_delayed}
	\end{minipage}
	
	\vspace{1em}
	
	\begin{minipage}[b]{0.31\textwidth}
		\centering
		\includegraphics[width=\textwidth]{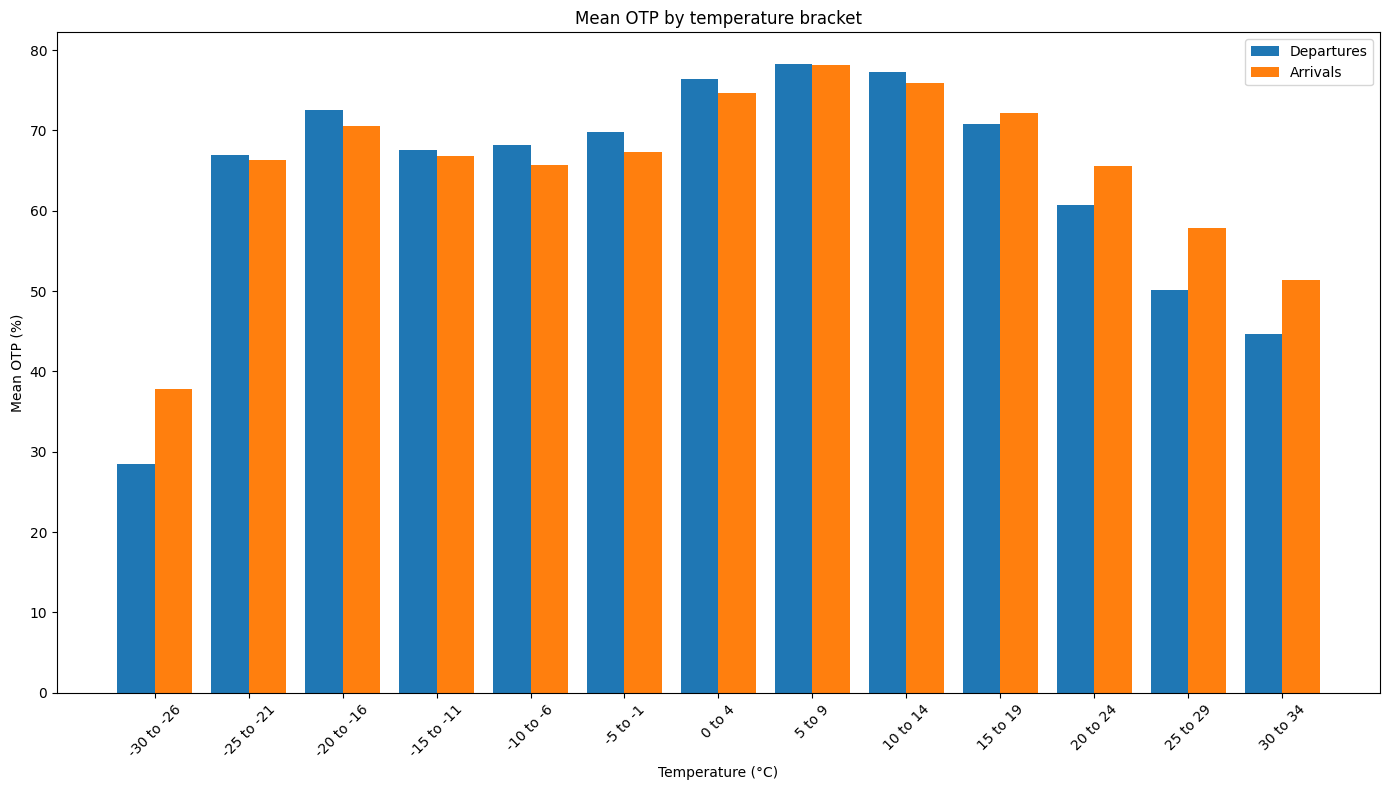}
		\caption{(d) OTP by temperature}
		\label{fig:qualitative_analysis_otp_temperature}
	\end{minipage}
	\hfill
	\begin{minipage}[b]{0.31\textwidth}
		\centering
		\includegraphics[width=\textwidth]{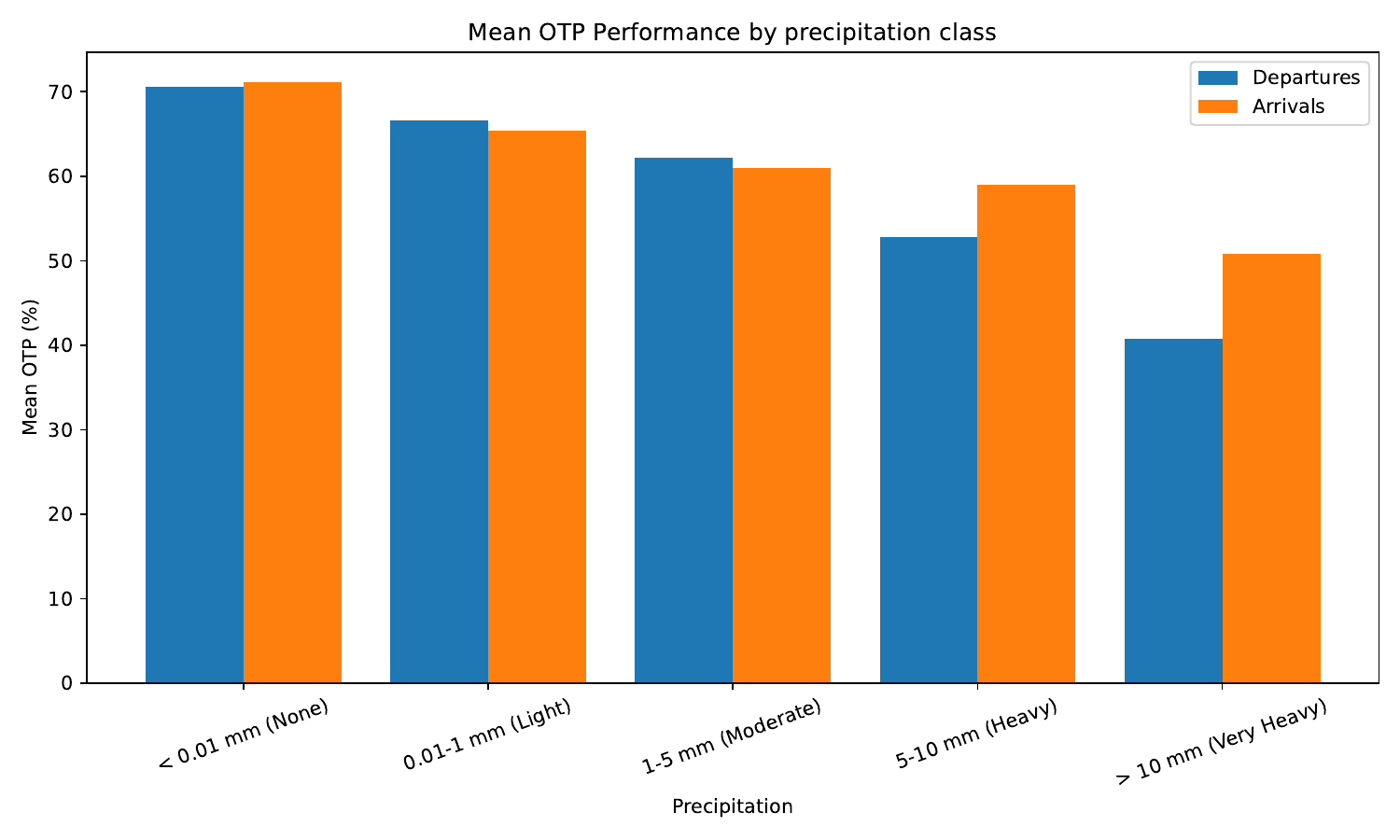}
		\caption{(e) OTP by precipitation level}
		\label{fig:qualitative_analysis_otp_precipitation}
	\end{minipage}
	\hfill
	\begin{minipage}[b]{0.31\textwidth}
		\centering
		\includegraphics[width=\textwidth]{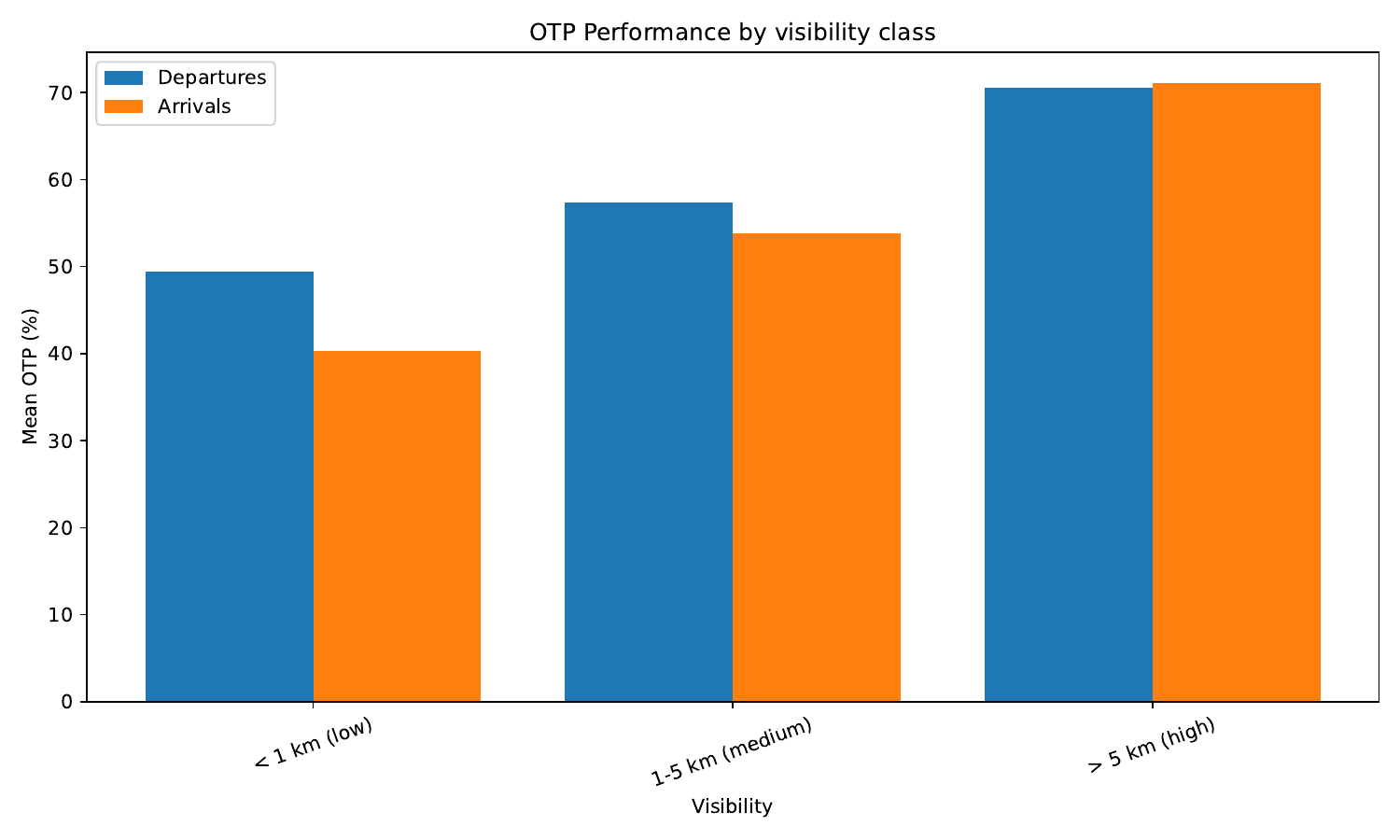}
		\caption{(f) OTP by visibility level}
		\label{fig:qualitative_analysis_otp_visibility}
	\end{minipage}
	
\end{figure}

We now examine how key features influence the likelihood of a flight being delayed. First, we observe that the overall performance of the airport tends to be lower during the summer and winter months. This pattern can be attributed to higher traffic volumes during peak vacation periods in July–August and December, as well as adverse weather conditions in the winter, such as snow and ice, which disrupt airport operations. Summer thunderstorms and winter storms also contribute significantly to increased delays. These seasonal effects are clearly visible in Figure~\ref{fig:qualitative_analysis_otp_month}, where OTP dips during July–August and December–February.

\begin{figure}[htbp]
	\centering
	\begin{minipage}[b]{0.48\textwidth}
		\centering
		\includegraphics[width=\textwidth]{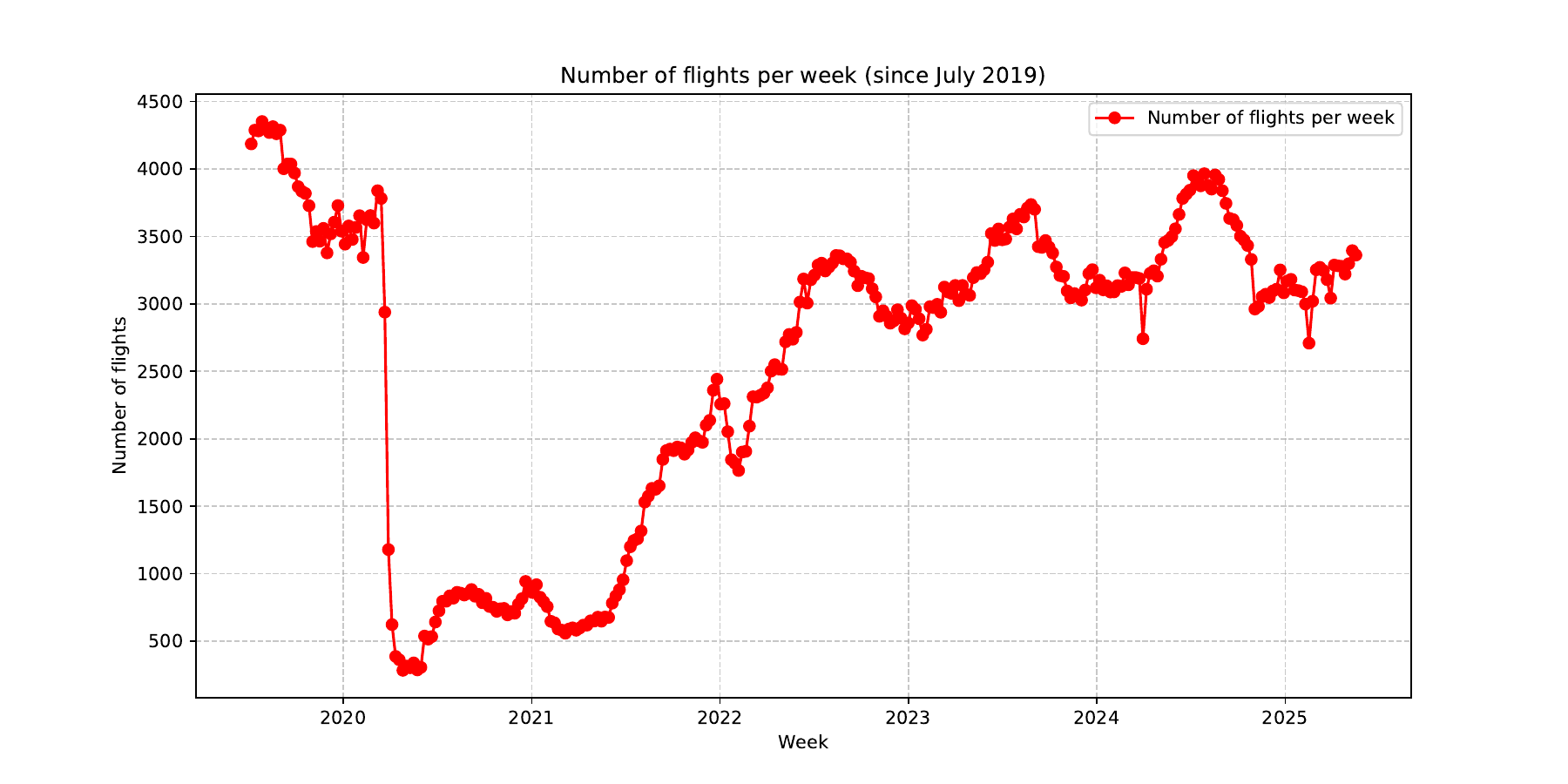}
		\caption{Number of flights per week since July 2019}
		\label{fig:nbFlights_wrt_time}
	\end{minipage}
	\hfill
	\begin{minipage}[b]{0.48\textwidth}
		\centering
		\includegraphics[width=\textwidth]{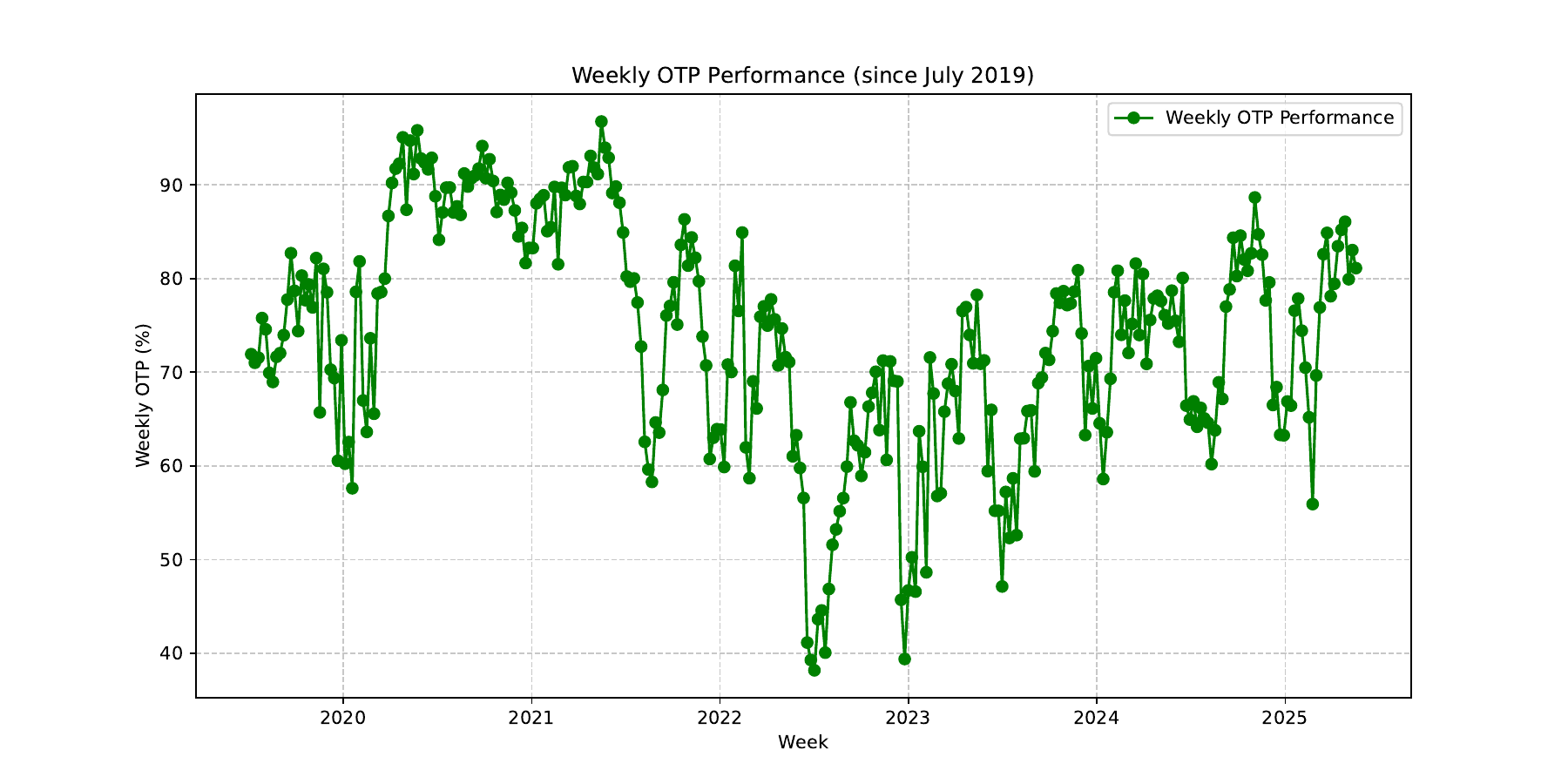}
		\caption{Weekly OTP since July 2019}
		\label{fig:weekly_otp}
	\end{minipage}
\end{figure}

Temporal effects are also evident at the daily level. Figure~\ref{fig:qualitative_analysis_otp_hour} shows OTP by hour of the day, highlighting a pronounced dip during the late afternoon and evening peak hours, which coincide with periods of highest airport activity. This suggests that congestion and resource constraints at peak times increase the likelihood of delays, in line with findings in prior studies on airport capacity \citep{ReboBala14}.

The impact of congestion can also be observed at the gate level: when one flight arrives or departs late, the subsequent user of that gate is more likely to miss the OTP threshold, as shown in Figure~\ref{fig:qualitative_analysis_otp_last_delayed}. This effect is consistent with the concept of delay propagation in airport operations \citep{BeatHsuBerrRome99, PyrgMaloOdon13}.

Weather-related effects further reinforce these dynamics. Figure~\ref{fig:qualitative_analysis_otp_temperature} demonstrates that extreme temperature conditions significantly reduce airport performance. Similarly, Figure~\ref{fig:qualitative_analysis_otp_precipitation} and Figure~\ref{fig:qualitative_analysis_otp_visibility} show that high levels of precipitation and low visibility both lead to marked drops in OTP. These weather conditions can create operational challenges, such as reduced runway capacity, increased taxi times, and the need for additional safety measures, all of which contribute to delays.

\subsection{Feature Engineering}

We enrich the set of variables used in the qualitative analysis to develop our models, defining a flight delay as in Definition~\ref{def:delay}.
For negative delays (early departures or arrivals), we set the delay duration to $0$. 

We use the K-means algorithm with $K=3$ to cluster airline sizes into three categories. The same algorithm is applied to cluster airport distance from the main airport into three categories. We introduce a binary variable indicating whether or not the flight is transatlantic. One-hot encoding is used for airline alliances, the section of the airport used by the flight (domestic, transborder, international), the season, and whether or not the flight occurs during the weekend (defined as Friday evening to Sunday evening). 

We also include a binary variable indicating whether the origin or destination airport is considered a MegaHub, following the classification proposed by \cite{OAG24}. The time of day is categorized as morning ($4 \leq \text{hour} < 10$), midday ($10 \leq \text{hour} < 16$), and evening ($16 \leq \text{hour} < 22$). We add a binary variable for whether or not the flight uses a remote gate. Finally, we include the number of gates and the number of flights appearing in the data within the same hour as the flight being predicted.

We add, as a feature, the predicted daily overall OTP of the airport. Beginning with the second week (the first week serving as the base dataset), we train a random forest model to predict the daily OTP of the airport. The features for this model include the season, the number of scheduled arrivals and departures for that day, the \textit{isWeekend} binary variable, seasonal binary indicators, and the average weather forecast for the day (averaged across all hours). In addition, we incorporate autoregressive features by including the OTP of the previous seven days. We then use this model to predict the OTP for the following week and add this predicted OTP as a feature to every flight scheduled in that week. This process is repeated iteratively until the OTP has been predicted for every day in the dataset.

For each gate, airline, and flight number in the dataset, we compute the historical OTP and the average delay duration of that entity, using only data prior to the flight being predicted. These variables are added as additional features.


Finally, we divide the dataset into a training set (80\%), a validation set (10\%), and a test set (10\%), and sample those sets randomly. 

\subsection{Modeling}

We begin by predicting whether a flight will be delayed or not. For this classification task, we employ multiple models: logistic regression, gradient boosting, and random forests. Hyperparameters are tuned using cross-validation, and model performance is evaluated using the $F_1$-score. The best-performing model is then selected. To assess feature relevance, we apply permutation feature importance on the test set. Features with importance values $>0$ are retained (since features with a negative impact yield importance $<0$ under permutation importance), and the model is retrained using only the retained features. The final model is then evaluated on the test set.

Next, we predict the delay duration of delayed flights. For this regression task, we test several models: linear regression, random forests, and gradient boosting. As with the classification task, hyperparameters are tuned via cross-validation. Model performance is primarily assessed using the MAE~\eqref{eq:MAE}, though we also consider the RMSE~\eqref{eq:RMSE} to ensure robustness. The model with the lowest MAE is selected; in cases where two models have comparable MAE, we select the model with the lowest RMSE.

After model selection, we again apply permutation feature importance to identify the most relevant features. Only the most important features are retained, and the model is retrained accordingly before final evaluation on the test set. In both the classification and regression tasks, the retained features are reported in the feature importance plots. The appendix presents the full list of initial features prior to feature selection. For interpretability, some features are added or removed in blocks; for example, airline alliance variables are either all retained or all removed.

This process is repeated separately for arrivals and departures.  
All experiments are implemented using the \textit{scikit-learn} library in Python \citep{Pedretal11}.

\section{Results and Discussion}
\label{sec:results}

We now present the main results obtained for both inference models and prediction models, considering departures and arrivals separately. 

\subsection{Inference Models}

The base features (prior to feature selection) used in the inference models are summarized in Table~\ref{tab:features_inference}.

For the classification task on departures, the best-performing model is gradient boosting. Its performance is reported in Table~\ref{tab:inference_class_perf}, with a weighted average $F_1$-score of 0.73, closely followed by the random forest model (0.72). These results on departure delays are consistent with findings from other studies on departure delay prediction~\citep{KhanMaChunWen21}.

\begin{figure}[htbp]
    \centering
    \begin{minipage}[b]{0.48\textwidth}
        \centering
        \includegraphics[width=\textwidth]{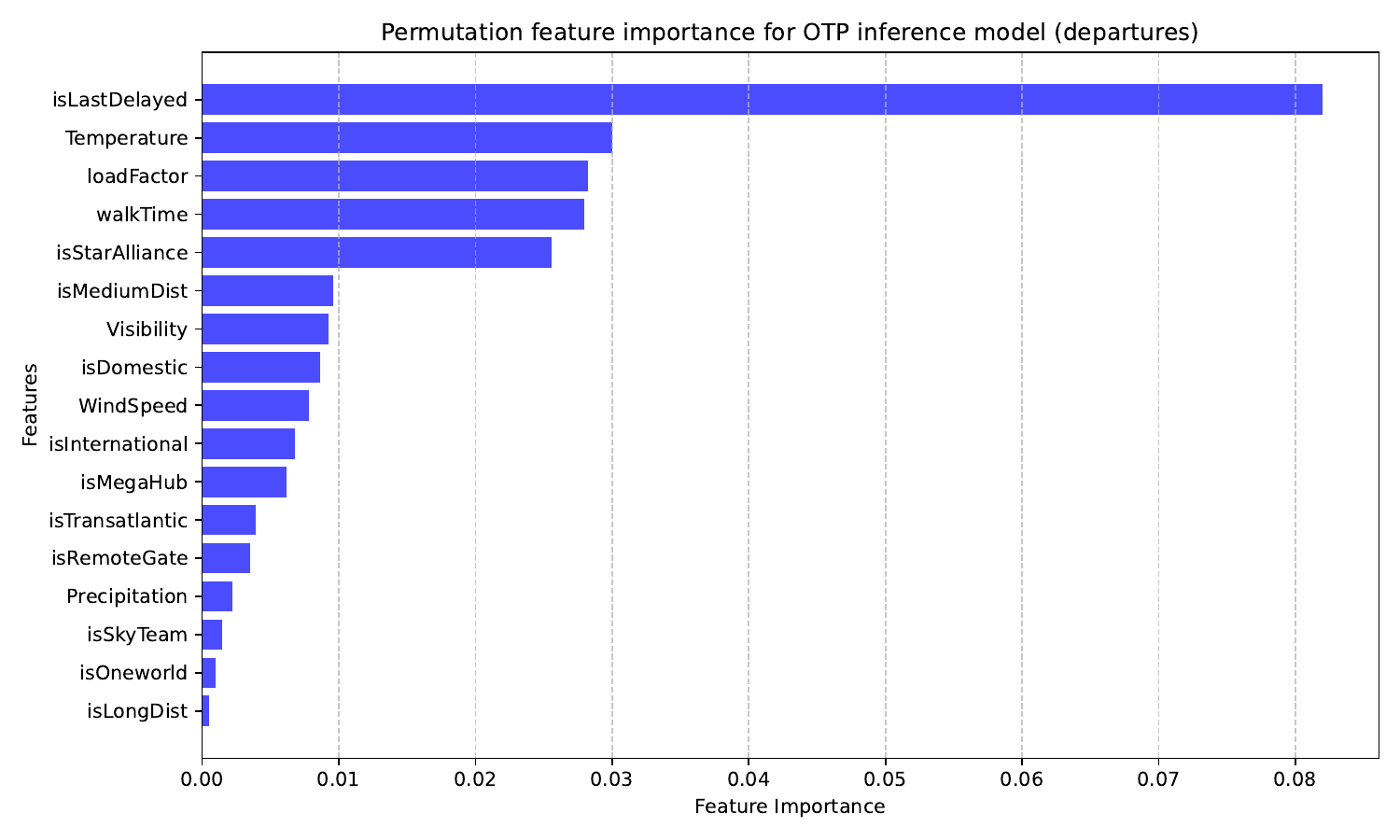}
        \caption{Feature importance for the inference classification model for departures}
        \label{fig:inference_class_feat_imp}
    \end{minipage}
    \hfill
    \begin{minipage}[b]{0.48\textwidth}
        \centering
        \includegraphics[width=\textwidth]{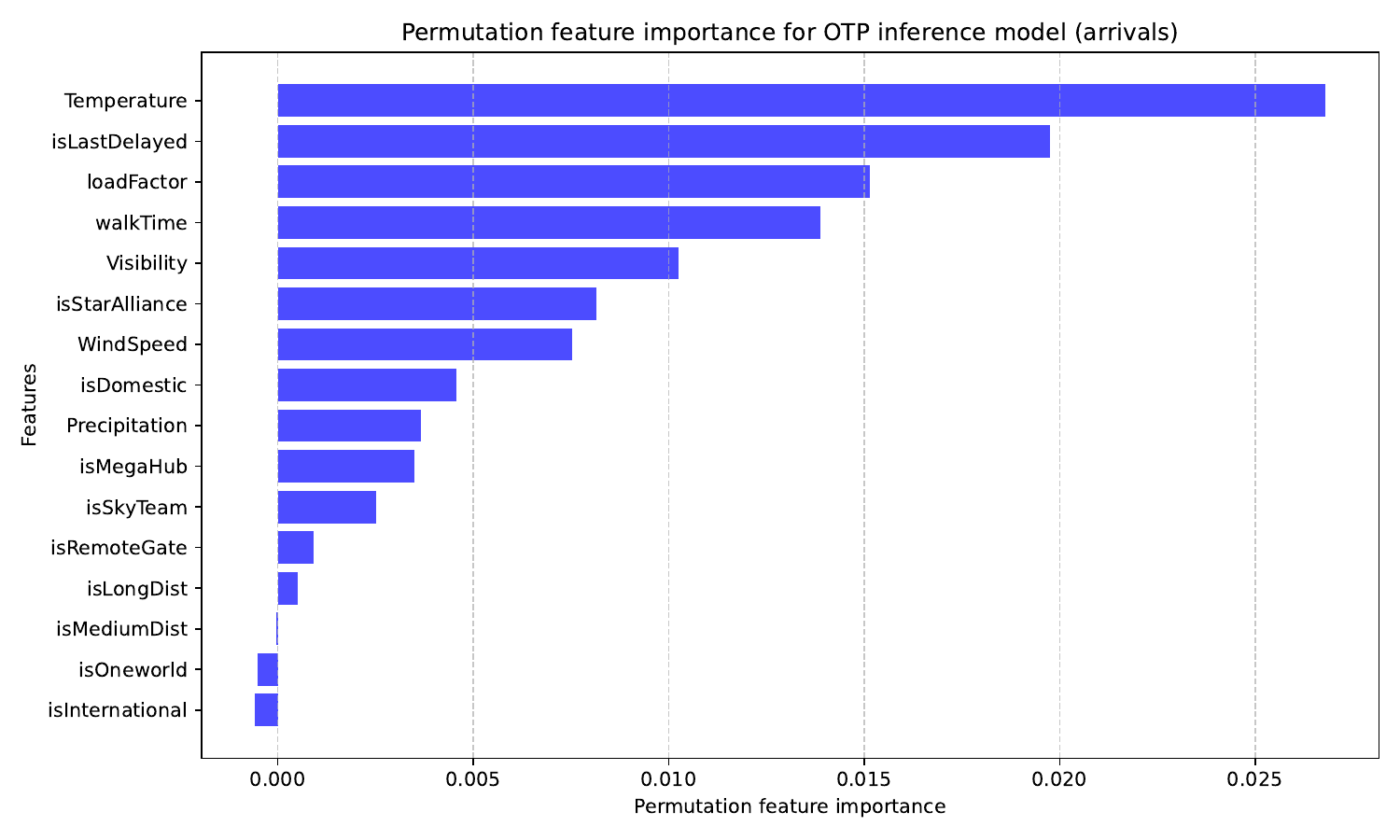}
        \caption{Feature importance for the inference classification model for arrivals}
        \label{fig:inference_class_feat_imp_arrivals}
    \end{minipage}
\end{figure}

From the feature importance plot in Figure~\ref{fig:inference_class_feat_imp}, we can see that the most important features are whether the last flight to use the gate was delayed and the walk time from security. This is particularly interesting because the first feature highlights a key opportunity to improve airport performance: by optimizing gate allocation and reducing aircraft turnaround times. The walk time from security being an important feature is also noteworthy, as its direct impact on performance is not immediately obvious. This suggests that walk time provides the model with information about the likelihood of a flight being delayed, even if it does not directly affect OTP. Further research is warranted in this area, but one plausible hypothesis is that flights with shorter walk times may be prioritized by the airport, for example through preferential gate allocation, which indirectly improves their on-time performance.

The performance of the classification model for arrivals is lower, with the best weighted average $F_1$-score of 0.68, again achieved by gradient boosting, with random forest as a close contender (0.67). The results are reported in Table~\ref{tab:inference_class_perf_arrivals}, and the feature importance plot is shown in Figure~\ref{fig:inference_class_feat_imp_arrivals}. 

We expect the performance of the arrival model to be lower, as arrivals are more dependent on external factors (e.g., weather or delays at the origin airport) that are not included in our dataset. The importance of the ``isLastDelayed'' feature decreases significantly, which is expected, since arrivals are less dependent on airport operations than departures.

\begin{table}[htbp]
	\caption{Performance of the inference model for departures.}
	\begin{tabular*}{\hsize}{@{\extracolsep{\fill}}llllll@{}}
		\toprule
		Model & Class & Precision & Recall & $F_1$-score & Observations\\
		\colrule
		Random Forest & Not delayed & 0.82 & 0.76 & 0.79 & 23781 \\
		& Delayed & 0.52 & 0.61 & 0.56 & 10270 \\
		\colrule
		& Macro average & 0.67 & 0.69 & 0.68 & 34051 \\
		& Weighted average & 0.73 & 0.71 & 0.72 & 34051 \\
		\colrule \colrule
		Logistic Regression & Not delayed & 0.73 & 0.94 & 0.82 & 23781 \\
		& Delayed & 0.60 & 0.21 & 0.32 & 10270 \\
		\colrule
		& Macro average & 0.67 & 0.58 & 0.57 & 34051 \\
		& Weighted average & 0.70 & 0.72 & 0.67 & 34051 \\
		\colrule \colrule
		Gradient Boosting & Not delayed & 0.77 & 0.92 & 0.84 & 23781 \\
		& Delayed & 0.66 & 0.36 & 0.47 & 10270 \\
		\colrule
		& Macro average & 0.72 & 0.64 & 0.65 & 34051 \\
		& Weighted average & 0.75 & 0.75 & 0.73 & 34051 \\
		\botrule
	\end{tabular*}
	\label{tab:inference_class_perf}
\end{table}

\begin{table}[htbp]
\caption{Performance of the inference model for arrivals.}
\begin{tabular*}{\hsize}{@{\extracolsep{\fill}}llllll@{}}
\toprule
Model & Class & Precision & Recall & $F_1$-score & Observations\\
\colrule
Random forest & Not delayed & 0.79 & 0.71 & 0.75 & 22013 \\
& Delayed & 0.44 & 0.55 & 0.49 & 9215 \\
\colrule
& Macro average & 0.62 & 0.63 & 0.62 & 31228 \\
& Weighted average & 0.69 & 0.66 & 0.67 & 31228 \\
\colrule \colrule
Logistic regression & Not delayed & 0.71 & 0.99 & 0.83 & 22013 \\
& Delayed & 0.56 & 0.02 & 0.03 & 9215 \\
\colrule
& Macro average & 0.63 & 0.51 & 0.43 & 31228 \\
& Weighted average & 0.66 & 0.71 & 0.59 & 31228 \\
\colrule \colrule
Gradient boosting & Not delayed & 0.74 & 0.94 & 0.83 & 22013 \\
& Delayed & 0.61 & 0.22 & 0.33 & 9215 \\
\colrule
& Macro average & 0.68 & 0.58 & 0.58 & 31228 \\
& Weighted average & 0.70 & 0.73 & 0.68 & 31228 \\
\botrule
\end{tabular*}
\label{tab:inference_class_perf_arrivals}
\end{table}

\begin{figure}[htbp]
    \centering
    \begin{minipage}[b]{0.48\textwidth}
        \centering
        \includegraphics[width=\textwidth]{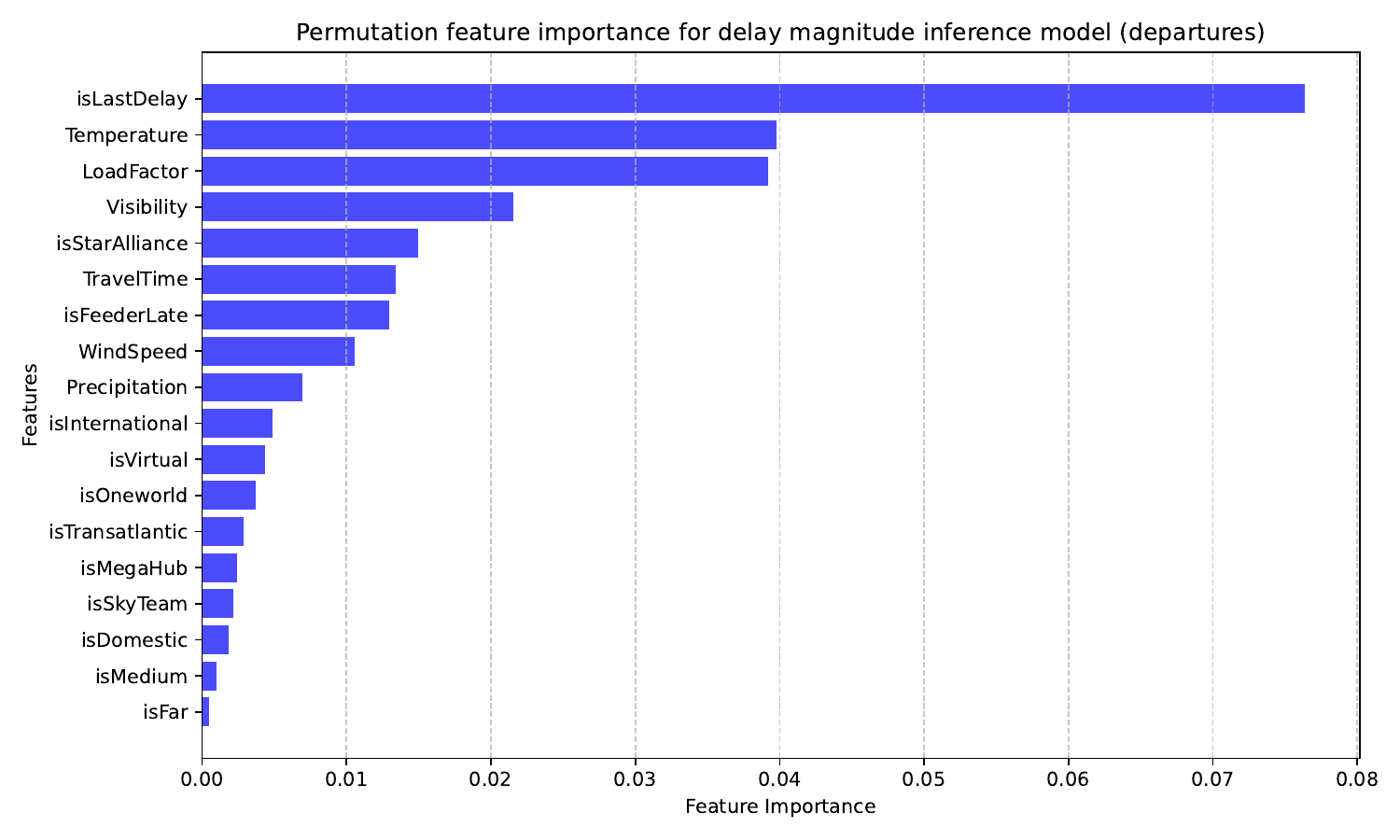}
        \caption{Feature importance for the inference regression model for departures}
        \label{fig:inference_reg_feat_imp}
    \end{minipage}
    \hfill
    \begin{minipage}[b]{0.48\textwidth}
        \centering
        \includegraphics[width=\textwidth]{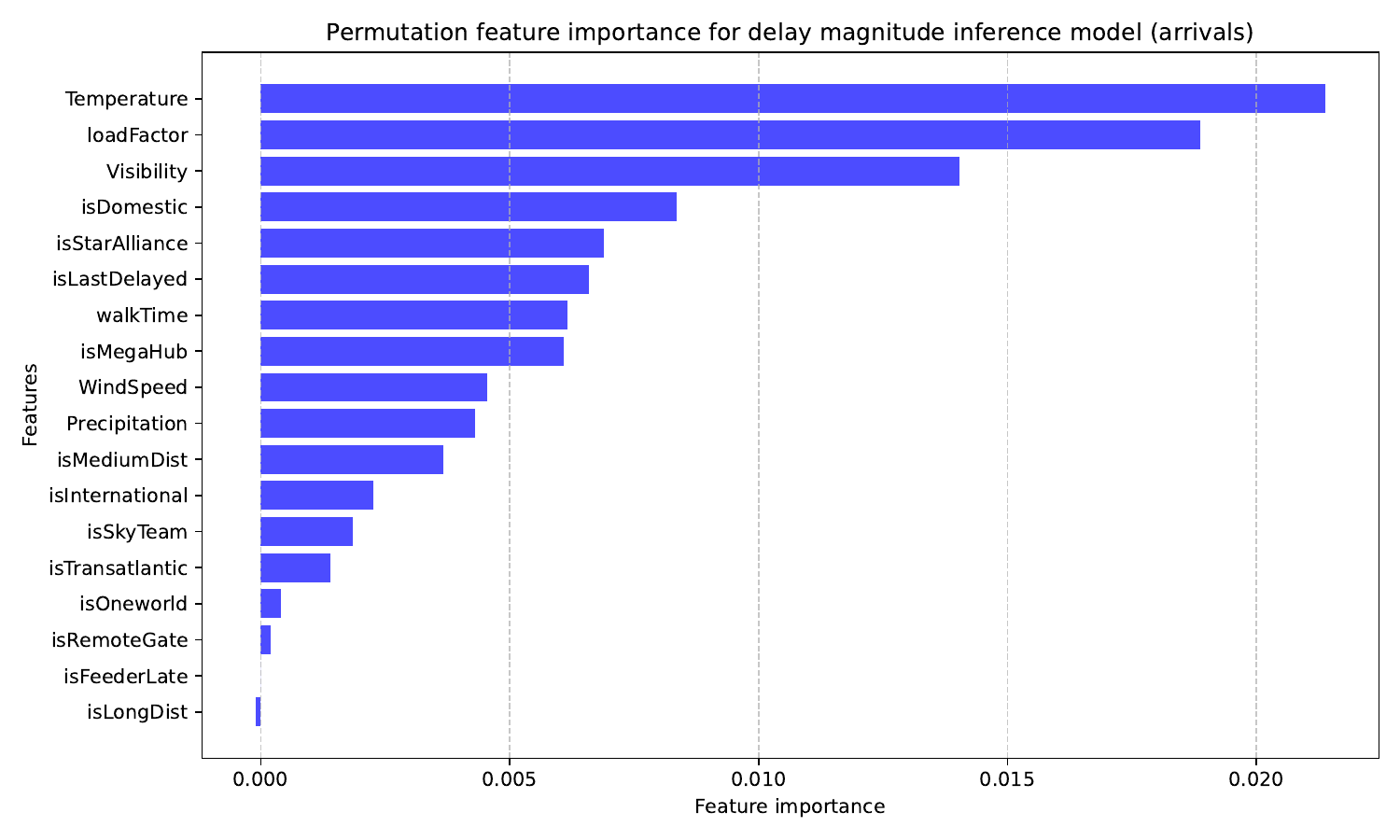}
        \caption{Feature importance for the inference regression model for arrivals}
        \label{fig:inference_reg_feat_imp_arrivals}
    \end{minipage}
\end{figure}

For the regression task, the best-performing model is gradient boosting. To account for the highly skewed distribution of delay durations (with a few extremely large delays), we train the model using median error instead of the usual mean squared error. We also experimented with training the models on log-transformed delays, but the results were less satisfactory. The performance of the model is reported in Table~\ref{tab:inference_reg_perf}, and the feature importance plot is shown in Figure~\ref{fig:inference_reg_feat_imp}. We observe that the most important features remain relatively unchanged compared to the OTP classification model.


\begin{table}[htbp]
\caption{Performance of the inference regression model for departures.}
\begin{tabular*}{\hsize}{@{\extracolsep{\fill}}lll@{}}
\toprule
Method & MAE & RMSE\\
\colrule
Gradient boosting & 17.31 minutes & 45.00 minutes\\
Linear regression & 22.55 minutes & 45.47 minutes\\
Random forest & 20.96 minutes & 43.80 minutes\\
\botrule
\end{tabular*}
\label{tab:inference_reg_perf}
\end{table}

The performance of the regression model for arrivals is noticeably lower, as shown in Table~\ref{tab:inference_reg_perf_arrivals}. This is consistent with the classification task, since arrivals are more dependent on external factors (such as weather or delays at the origin airport) that are not available in our data. The importance of the ``isLastDelayed'' feature once again decreases significantly, which is expected, as arrivals are less dependent on airport operations than departures.

\begin{table}[htbp]
\caption{Performance of the inference regression model for arrivals.}
\begin{tabular*}{\hsize}{@{\extracolsep{\fill}}lll@{}}
\toprule
Model & MAE & RMSE\\
\colrule
Gradient boosting & 20.63 minutes & 73.54 minutes\\
Linear regression & 27.10 minutes & 73.25 minutes\\
Random forest & 25.71 minutes & 69.68 minutes\\
\botrule
\end{tabular*}
\label{tab:inference_reg_perf_arrivals}
\end{table}

\subsection{Prediction Models}

The set of features (prior to selection) is presented in Table~\ref{tab:features_prediction}.


The performance of the classification models for departures is reported in Table~\ref{tab:prediction_class_perf}. Both the random forest and gradient boosting models achieve a weighted average $F_1$-score of 0.71, indicating reasonably good performance.

\begin{table}[htbp]
\caption{Performance of the prediction classification model for departures.}
\begin{tabular*}{\hsize}{@{\extracolsep{\fill}}llllll@{}}
\toprule
Model & Class & Precision & Recall & $F_1$-score & Support\\
\colrule
Random forest & Not delayed & 0.78 & 0.74 & 0.76 & 10647 \\
&Delayed  & 0.58 & 0.64 & 0.61 & 5986 \\
\colrule
& Macro average & 0.68 & 0.69 & 0.69 & 16633 \\
& Weighted average & 0.71 & 0.70 & 0.71 & 16633 \\
\colrule \colrule
Logistic regression & Not delayed & 0.72 & 0.87 & 0.79 & 10647 \\
& Delayed & 0.62 & 0.39 & 0.48 & 5986 \\
\colrule
& Macro average & 0.67 & 0.63 & 0.63 & 16633 \\
& Weighted average & 0.68 & 0.70 & 0.68 & 16633 \\
\colrule \colrule
Gradient boosting & Not delayed & 0.74 & 0.87 & 0.80 & 10647 \\
& Delayed & 0.66 & 0.46 & 0.54 & 5986 \\
\colrule
& Macro average & 0.70 & 0.66 & 0.67 & 16633 \\
& Weighted average & 0.71 & 0.72 & 0.71 & 16633 \\
\botrule
\end{tabular*}
\label{tab:prediction_class_perf}
\end{table}

\begin{figure}[htbp]
	\centering
	\begin{minipage}[b]{0.48\textwidth}
		\centering
		\includegraphics[width=\textwidth]{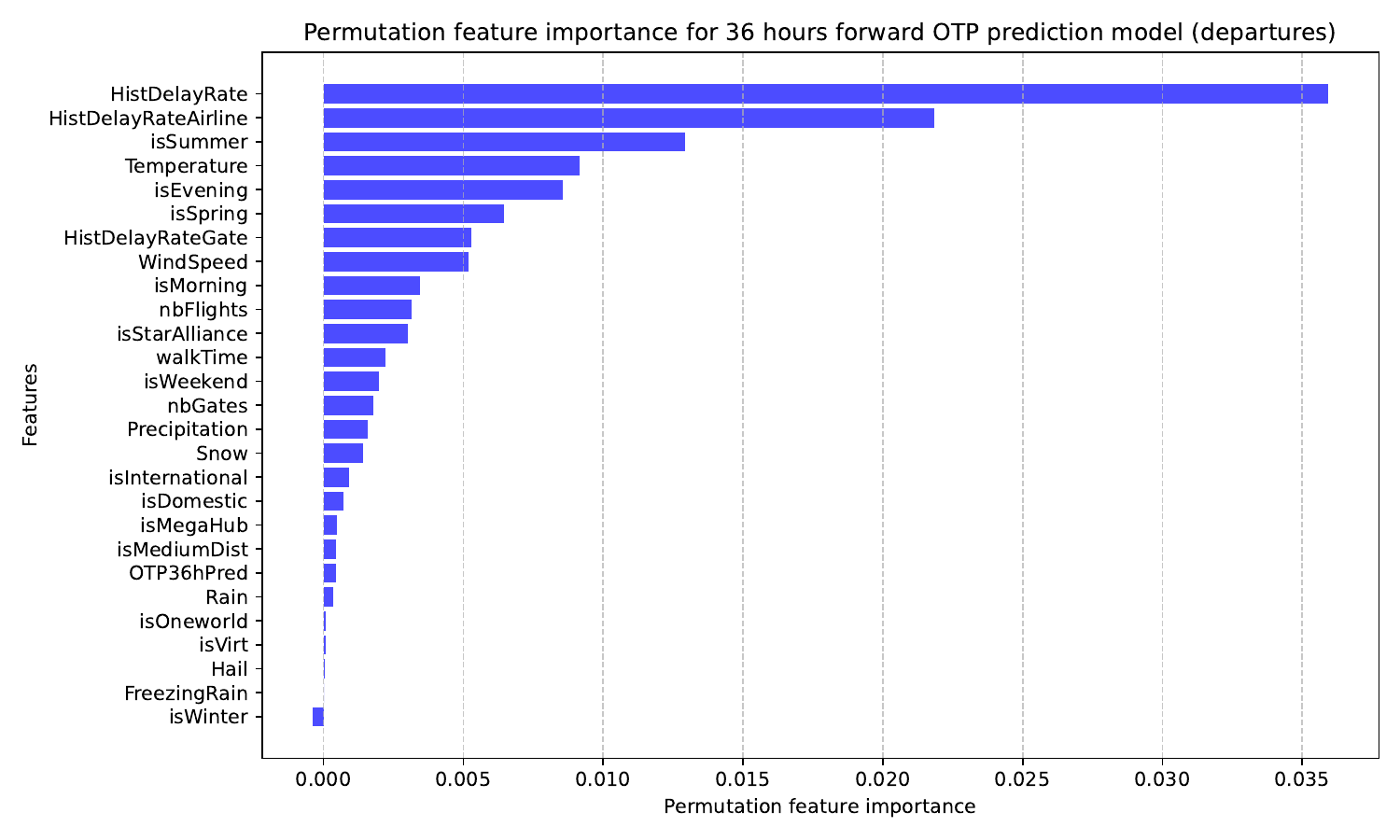}
		\caption{Feature importance for the prediction classification model for departures}
		\label{fig:prediction_class_feat_imp}
	\end{minipage}
	\hfill
	\begin{minipage}[b]{0.48\textwidth}
		\centering
		\includegraphics[width=\textwidth]{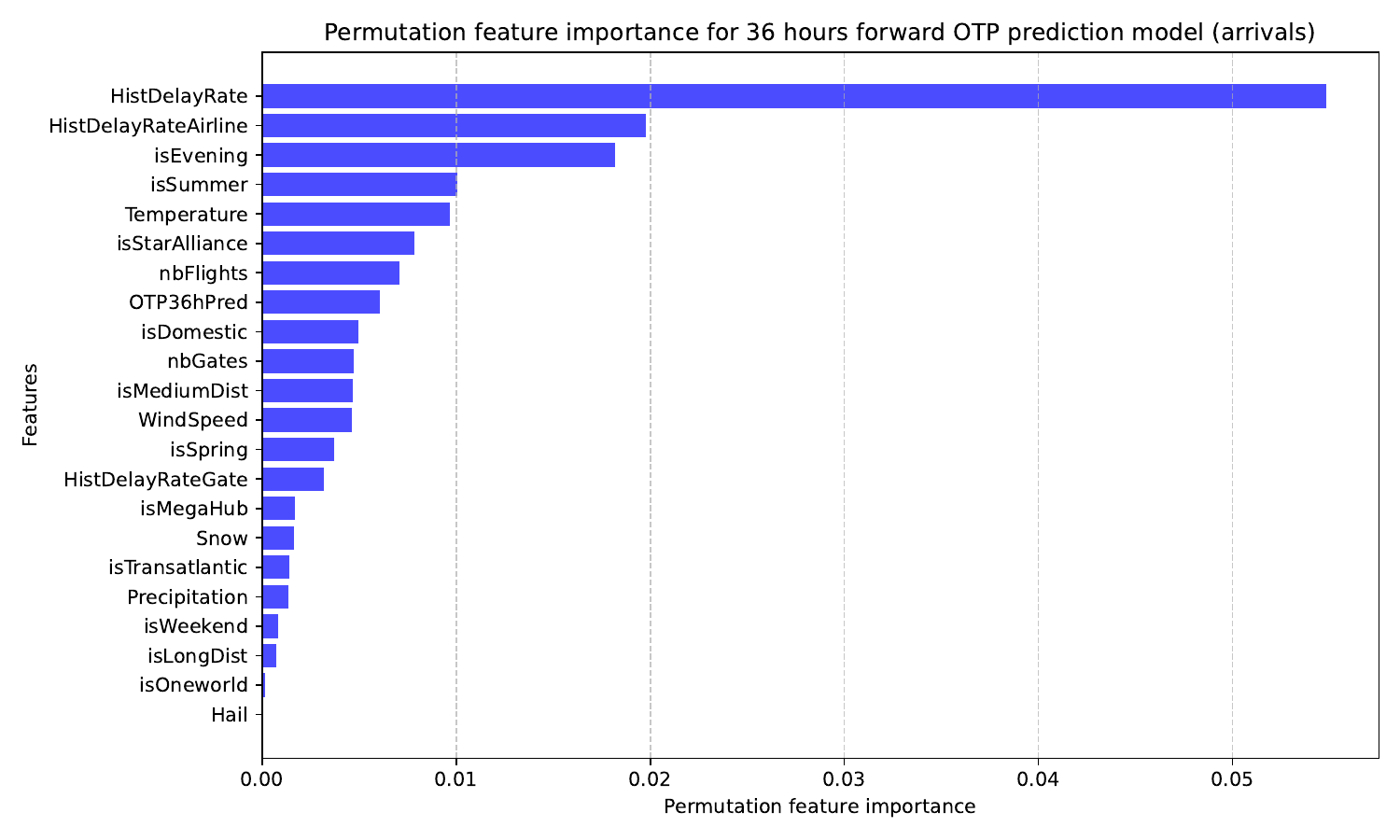}
		\caption{Feature importance for the prediction classification model for arrivals}
		\label{fig:prediction_class_feat_imp_arrivals}
	\end{minipage}
\end{figure}

The feature importance plot is shown in Figure~\ref{fig:prediction_class_feat_imp}. The most influential features are the historical delay rate of the flight number and the historical delay rate of the airline. This indicates that certain flight routes and airlines exhibit a higher propensity for delays, which may be attributed to underlying factors such as operational efficiency, route complexity, and resource allocation.

The performance of the classification model for arrivals is slightly lower, although the weighted average $F_{1}$-score remains at $0.70$ for both random forest and gradient boosting. The results are presented in Table~\ref{tab:prediction_class_perf_arrivals}, and the corresponding feature importance plot is shown in Figure~\ref{fig:prediction_class_feat_imp_arrivals}. The most influential features are consistent with those in the departure model, with the historical delay rates of the flight number and the airline emerging as the most important.

\begin{table}[htbp]
\caption{Performance of the prediction classification model for arrivals.}
\begin{tabular*}{\hsize}{@{\extracolsep{\fill}}llllll@{}}
\toprule
Model & Class & Precision & Recall & $F_1$-score & Support\\
\colrule
Random forest & Not delayed & 0.79 & 0.74 & 0.76 & 9880 \\
& Delayed  & 0.54 & 0.62 & 0.58 & 5025 \\
\colrule
& Macro average & 0.67 & 0.68 & 0.67 & 14905 \\
& Weighted average & 0.71 & 0.70 & 0.70 & 14905 \\
\colrule \colrule
Logistic regression & Not delayed & 0.71 & 0.91 & 0.80 & 9880 \\
& Delayed & 0.61 & 0.27 & 0.38 & 5025 \\
\colrule
& Macro average & 0.66 & 0.59 & 0.59 & 14905 \\
& Weighted average & 0.68 & 0.70 & 0.66 & 14905 \\
\colrule \colrule
Gradient boosting & Not delayed & 0.74 & 0.89 & 0.81 & 9880 \\
& Delayed & 0.64 & 0.39 & 0.49 & 5025 \\
\colrule
& Macro average & 0.69 & 0.64 & 0.65 & 14905 \\
& Weighted average & 0.71 & 0.72 & 0.70 & 14905 \\
\botrule
\end{tabular*}
\label{tab:prediction_class_perf_arrivals}
\end{table}

\begin{figure}[htbp]
	\centering
	\begin{minipage}[b]{0.48\textwidth}
		\centering
		\includegraphics[width=\textwidth]{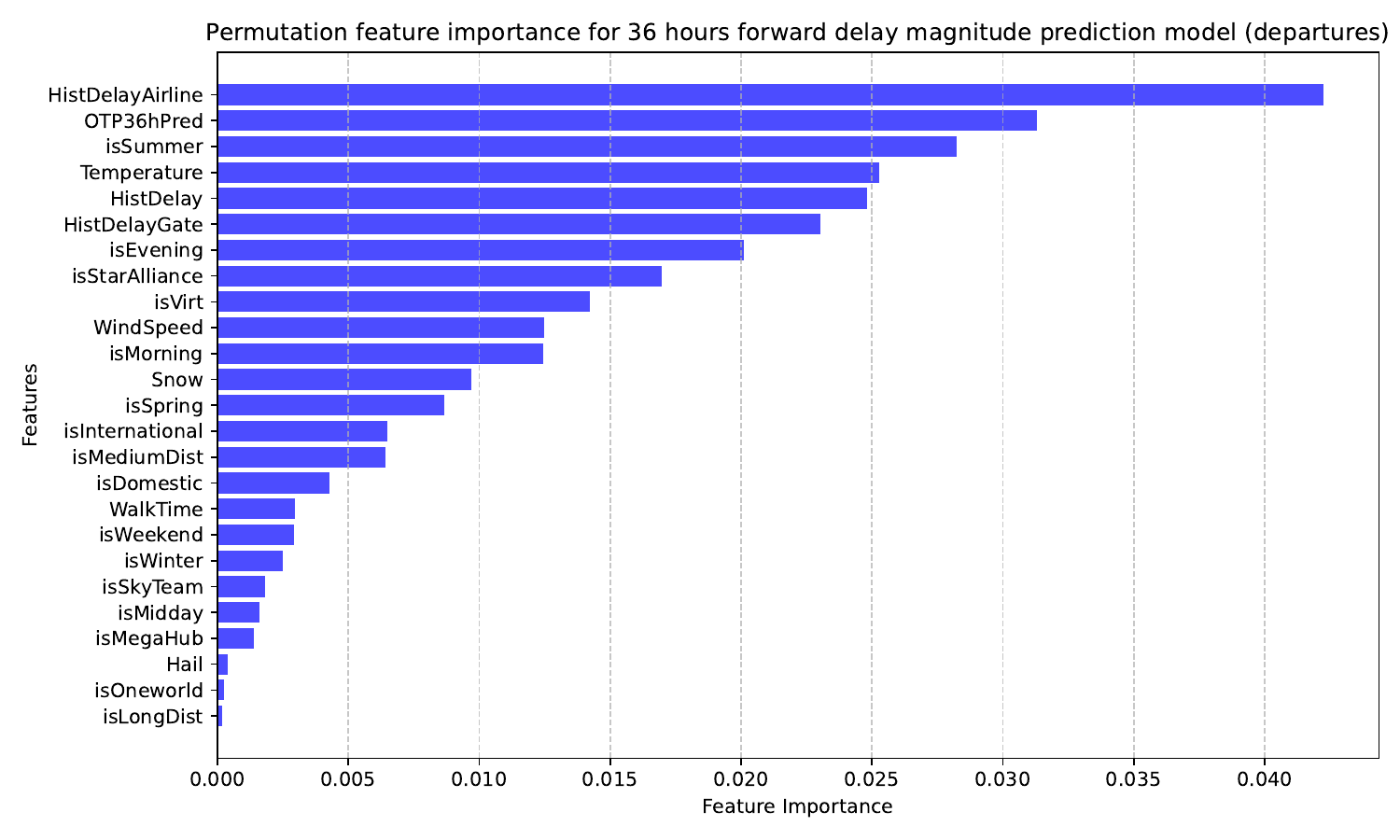}
		\caption{Feature importance for the prediction regression model for departures}
		\label{fig:prediction_reg_feat_imp}
	\end{minipage}
	\hfill
	\begin{minipage}[b]{0.48\textwidth}
		\centering
		\includegraphics[width=\textwidth]{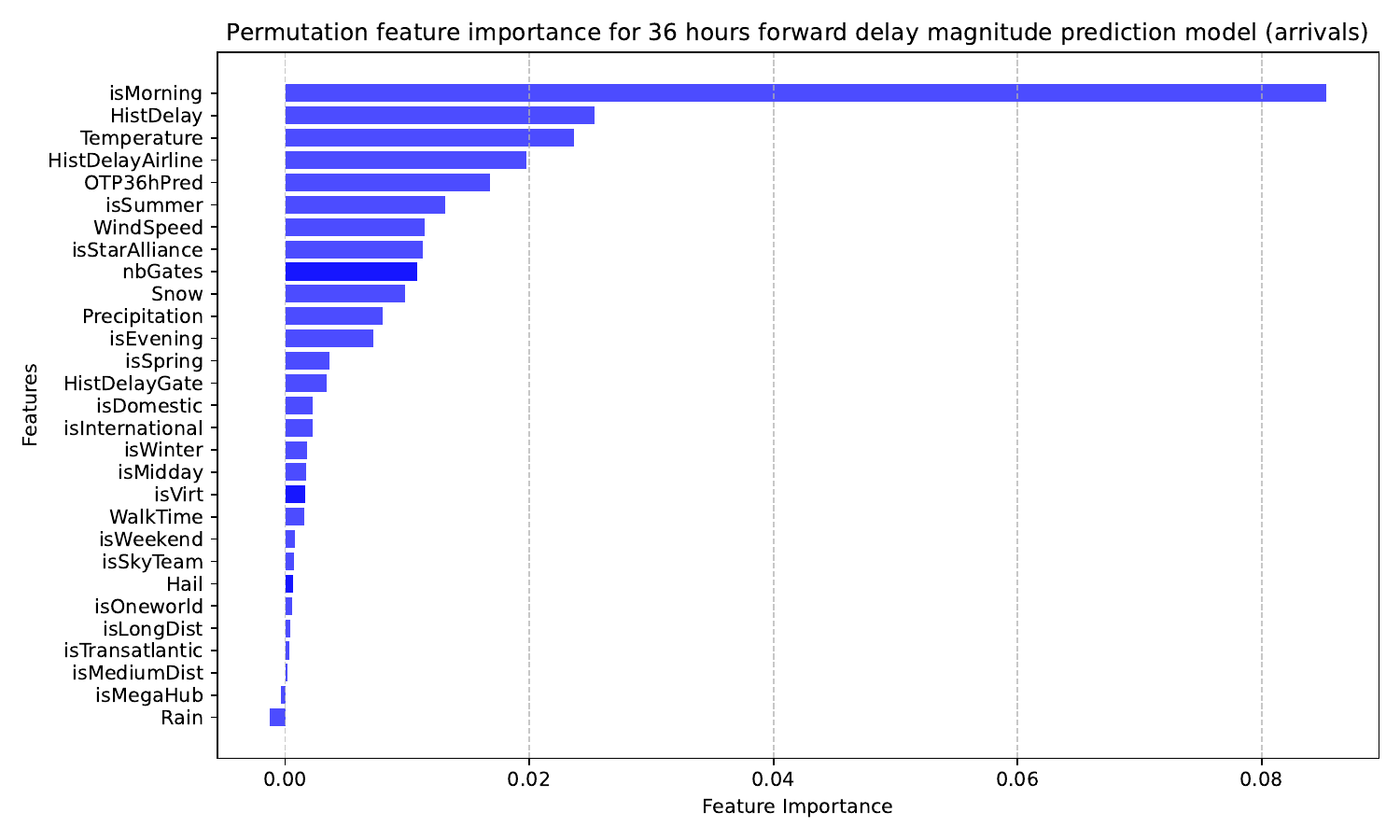}
		\caption{Feature importance for the prediction regression model for arrivals}
		\label{fig:prediction_reg_feat_imp_arrivals}
	\end{minipage}
\end{figure}

The performance of the regression models for departures is reported in Table~\ref{tab:prediction_reg_perf}, with the corresponding feature importance plots shown in Figure~\ref{fig:prediction_reg_feat_imp}. The regression model for arrivals performs only slightly worse, as shown in Table~\ref{tab:prediction_reg_perf_arrivals}, and its feature importance plot is presented in Figure~\ref{fig:prediction_reg_feat_imp_arrivals}. The overall feature importance remains relatively similar, with the notable exception of the ``isMorning'' feature, which becomes unexpectedly important for arrivals.

\begin{table}[htbp]
\caption{Performance of the prediction regression model for departures.}
\begin{tabular*}{\hsize}{@{\extracolsep{\fill}}lll@{}}
\toprule
Model & MAE & RMSE\\
\colrule
Gradient boosting & 20.11 minutes & 53.51 minutes\\
Linear regression & 25.07 minutes & 56.74 minutes\\
Random forest & 23.66 minutes & 54.54 minutes\\
\botrule
\end{tabular*}
\label{tab:prediction_reg_perf}
\end{table}

\begin{table}[htbp]
\caption{Performance of the prediction regression model for arrivals.}
\begin{tabular*}{\hsize}{@{\extracolsep{\fill}}lll@{}}
\toprule
Model & MAE & RMSE\\
\colrule
Gradient boosting & 20.82 minutes & 56.27 minutes\\
Linear regression & 26.09 minutes & 57.12 minutes\\
Random forest & 24.83 minutes & 55.59 minutes\\
\botrule
\end{tabular*}
\label{tab:prediction_reg_perf_arrivals}
\end{table}

\section{Application to airport operations}
\label{sec:application}

We developed the aforementioned models in close collaboration with a major international airport, which is interested in leveraging these models to improve its operations. To support this goal, we created several tools to assist in operational decision-making, relying on previously developed models.

\subsection{Flight-by-Flight Likelihood of Delay and Delay Duration}

Similarly to \citet{BeltRibedeWiSun25}, we developed a tool to estimate both the likelihood of a specific flight being delayed and the expected duration of the delay.

\begin{figure}[htbp]
	\centering
	\includegraphics[width=0.6\textwidth]{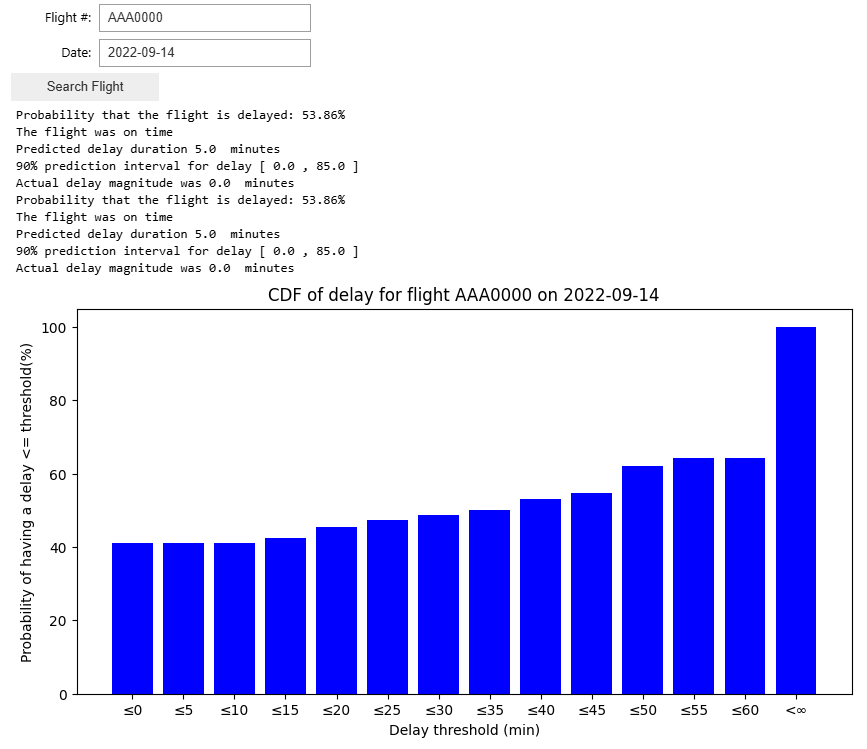}
	\caption{Example of delay duration distribution for a specific flight}
	\label{fig:delay_distribution_example}
\end{figure}

We begin by training a series of binary regression models to predict the probability that a delay exceeds a given threshold, with thresholds ranging from 0 to 60 minutes in increments of 5 minutes.
Any binary regression model can be used for this purpose; in our case, we employ random forests.
Each model outputs the probabilities associated with the two possible outcomes—delay exceeding or not exceeding the specified threshold.
The increment size is chosen to be sufficiently large to reduce the risk of overestimating the probability of longer delays.
For each flight and each threshold, the final probability of delay is defined as the minimum between the predicted probability and that of a delay exceeding the previous threshold.
This procedure yields a cumulative distribution function (CDF) for the delay duration of the flight. An example of such a distribution is shown in Figure~\ref{fig:delay_distribution_example}.



We also predict the expected delay duration of the flight using the regression model. Since we trained our models using quantile loss, it is straightforward to extract prediction intervals for the delay duration. Specifically, we used the 50th quantile (median) to train the regression model, but the 5th and 95th quantiles can also be used to obtain a 90\% prediction interval for the delay duration.

\subsection{Scenario Simulation Tool}

We also developed a scenario simulation tool to help the airport understand how different weather conditions can impact its performance over a season. 

The features used include the number of each type of disruptive event during the season, such as snowstorms, freezing rain, heavy rain, thunderstorms, and operational events (e.g., security issues). The tool also allows the user to input the average turnaround time allocated for flights over the season. This turnaround time is represented as a factor between 0.75 and 2, corresponding to the proportion of the optimal turnaround time allocated to each flight. For example, if a flight has an optimal turnaround time of 30 minutes and the user inputs a factor of 1.5, the model assumes a turnaround time of 45 minutes for that flight.

We separately compute the average performance loss or gain associated with each event, as illustrate in Figure~\ref{tab:weather_impact_detailed},
as well as the impact of turnaround time, which is estimated using a simple linear regression of historical OTP on planned turnaround time. A weighted sum of these losses and gains is then used to compute the overall performance change for the season. This overall adjustment is applied to the historical OTP of the airport to predict the seasonal OTP.
The predicted seasonal OTP is therefore computed as
$$
\left(\text{Average season OTP} + \sum_{\text{event}} n_\text{event} \cdot \text{performance gain/loss}_\text{event}\right) \cdot \text{performance gain/loss}_\text{turnaround},
$$
where $n_\text{event}$ is the number of occurrences of each event during the season, the performance gain or loss for events is the historical impact of each event (additive), and the performance gain or loss for turnaround time is the average impact of the allocated turnaround time (multiplicative).
We favor such a simple approach, as the limited amount of historical data on extreme weather and operational events renders more complex techniques susceptible to overfitting.

\begin{figure}[htbp]
	\centering
	\includegraphics[width=0.7\textwidth]{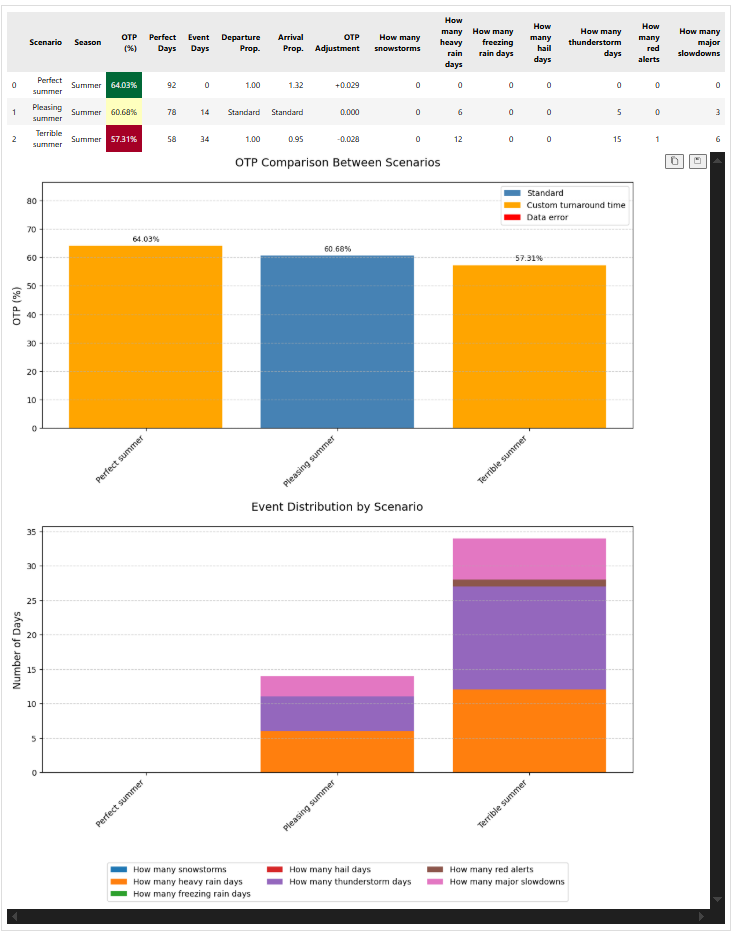}
	\caption{Example output of the scenario simulation tool}
	\label{fig:scenario_simulation_tool}
\end{figure}

This tool, illustrated in Figure~\ref{fig:scenario_simulation_tool}, remains relatively simple, with several opportunities for future development.
First, additional features could be modeled, such as operational events or staffing levels. Second, the impact of weather events could be modeled more accurately by incorporating their timing, duration, and intensity. Finally, interactions between different events could be considered, as their combined impact is unlikely to be purely additive. Given the relatively small sample of extreme weather and operational events, incorporating information from other airports would also likely improve the model's reliability.

\section{Conclusion}
\label{sec:conclusion}

In this study, we analyzed delays at a major international airport and developed models to predict both the likelihood and duration of these delays. We found that certain features, such as the historical delay rates of flight numbers and airlines, are strong predictors. Our models achieved reasonable performance, with a weighted average $F_1$-score of 0.71 for the classification of delayed and a mean absolute error of 20.11 minutes for the regression-based estimation of delays. These findings can help airlines and airports better manage operations and improve on-time performance.

We observed that weather conditions, as expected, play a significant role in flight delays. Additionally, other factors, such as runway traffic, walk time from security to gate, historical delays of the airline, gate, and flight route, as well as time of day and overall traffic, were found to be important predictors. The models we developed can forecast delays up to 36 hours in advance, though the methodology can be extended to other time horizons.

Several areas remain for future research. The impact of air traffic control and airport infrastructure on delays has not been fully explored. Further studies could also assess the effectiveness of interventions designed to reduce delays. Moreover, incorporating real-time data, weather forecasts from origin and destination airports, and air traffic control information may further improve predictive accuracy. The influence of walk time on delays also warrants additional investigation.

\section*{Acknowledgements}

This project was conducted under the MITACS grant IT43302, with the support of IVADO. The authors also acknowledge the collaboration of the airport partner, who provided access to the data used in this study.

\appendix

\section{Variables description}
\label{sec:appendix}

\begin{table}[htbp]
\caption{Features used for the inference models.}
\begin{tabular*}{\hsize}{@{\extracolsep{\fill}}lll@{}}
\toprule
Feature & Description & Type\\
\colrule
isStarAlliance & Indicates if the flight is operated by Star Alliance & Binary \\
isSkyTeam & Indicates if the flight is operated by SkyTeam & Binary \\
isOneworld & Indicates if the flight is operated by Oneworld & Binary \\
Visibility & Visibility at the time of departure/arrival (km) & Continuous \\
Precipitation & Amount of precipitation at the time of departure/arrival (mm) & Continuous \\
WindSpeed & Wind speed at the time of departure/arrival (km/h) & Continuous \\
Temperature & Temperature at the time of departure/arrival (°C) & Continuous \\
nbGates & Number of gates in use at the time of departure/arrival & Continuous \\
nbFlights & Number of flights scheduled to depart/arrive at the same hour & Continuous \\
isRemoteGate & Indicates if the flight is using a remote gate & Binary \\
isDomestic & Indicates if the flight is in the domestic section of the airport & Binary \\
isInternational & Indicates if the flight is in the international section of the airport & Binary \\
isLastDelayed & Indicates if the last flight to use the gate was delayed & Binary \\
walkTime & Walk time from security to gate (minutes) & Continuous \\
isMegaHub & Indicates if the origin/destination airport is a MegaHub & Binary \\
isTransatlantic & Indicates if the flight is transatlantic & Binary \\
loadFactor & Average monthly load factor of the flight's section of the airport & Continuous \\
isMediumDist & Indicates if the flight distance is medium (based on K-means clustering) & Binary \\
isLongDist & Indicates if the flight distance is long (based on K-means clustering) & Binary \\
isEU & Indicates if the flight is to/from Europe & Binary \\
isMENA & Indicates if the flight is to/from Middle East and North Africa & Binary \\
isAPAC & Indicates if the flight is to/from Asia-Pacific & Binary \\
isLAC & Indicates if the flight is to/from Latin America and Caribbean & Binary \\
\botrule
\end{tabular*}
\label{tab:features_inference}
\end{table}

\begin{table}[htbp]
\caption{Features used for the prediction models.}
\begin{tabular*}{\hsize}{@{\extracolsep{\fill}}lll@{}}
\toprule
Feature & Description & Type\\
\colrule
HistDelayRate & Historical delay rate of the flight number & Continuous \\
HistDelayRateGate & Historical delay rate of the gate & Continuous \\
HistDelayRateAirline & Historical delay rate of the airline & Continuous \\
OTP36hPred & Predicted OTP of the airport for the day of the flight & Continuous \\
isStarAlliance & Indicates if the flight is operated by Star Alliance & Binary \\
isSkyTeam & Indicates if the flight is operated by SkyTeam & Binary \\
isOneworld & Indicates if the flight is operated by Oneworld & Binary \\
Visibility & Visibility at the time of departure/arrival (km) & Continuous \\
Precipitation & Amount of precipitation at the time of departure/arrival (mm) & Continuous \\
WindSpeed & Wind speed at the time of departure/arrival (km/h) & Continuous \\
Temperature & Temperature at the time of departure/arrival (°C) & Continuous \\
Snow & Forecasted amount of snow (cm) & Continuous \\
FreezingRain & Forecasted amount of freezing rain (cm) & Continuous \\
IcePellets & Forecasted amount of ice pellets (cm) & Continuous \\
Rain & Forecasted amount of rain (cm) & Continuous \\
nbGates & Number of gates in use at the time of departure/arrival & Continuous \\
nbFlights & Number of flights scheduled to depart/arrive at the same hour & Continuous \\
isWinter & Indicates if the flight is in winter & Binary \\
isSpring & Indicates if the flight is in spring & Binary \\
isSummer & Indicates if the flight is in summer & Binary \\
isFall & Indicates if the flight is in fall & Binary \\
isWeekend & Indicates if the flight is on a weekend & Binary \\
isMorning & Indicates if the flight is in the morning & Binary \\
isMidday & Indicates if the flight is in midday & Binary \\
isEvening & Indicates if the flight is in the evening & Binary \\
isRemoteGate & Indicates if the flight is using a remote gate & Binary \\
isDomestic & Indicates if the flight is in the domestic section of the airport & Binary \\
isInternational & Indicates if the flight is in the international section of the airport & Binary \\
walkTime & Walk time from security to gate (minutes) & Continuous \\
isMegaHub & Indicates if the origin/destination airport is a MegaHub & Binary \\
isTransatlantic & Indicates if the flight is transatlantic & Binary \\
isMediumDist & Indicates if the flight distance is medium (based on K-means clustering) & Binary \\
isLongDist & Indicates if the flight distance is long (based on K-means clustering) & Binary \\
isEU & Indicates if the flight is to/from Europe & Binary \\
isMENA & Indicates if the flight is to/from Middle East and North Africa & Binary \\
isAPAC & Indicates if the flight is to/from Asia-Pacific & Binary \\
isLAC & Indicates if the flight is to/from Latin America and Caribbean & Binary \\
\botrule
\end{tabular*}
\label{tab:features_prediction}
\end{table}

\begin{table}[htbp]
\caption{Weather event impact analysis by season.}
\begin{tabular*}{\hsize}{@{\extracolsep{\fill}}llrrrrr@{}}
\toprule
Season & Event & Hours & Days & OTP (\%) & Cancel (\%)\\
\colrule
Summer & Baseline & --- & --- & 55.82 & 5.48 \\
& Heavy Rain & 115 & 60 & 49.66 & 8.38\\
\colrule
Fall & Baseline & --- & --- & 70.47 & 3.93\\
& Heavy Rain & 63 & 24 & 66.31 & 5.34\\
& Snowstorm & 19 & 4 & 54.65 & 5.76\\
& Freezing Rain & 5 & 2 & 77.05 & 6.76\\
\colrule
Winter & Baseline & --- & --- & 61.77 & 6.79\\
& Heavy Rain & 16 & 7 & 35.89 & 25.94\\
& Snowstorm & 186 & 35 & 36.20 & 16.46\\
& Freezing Rain & 15 & 9 & 61.68 & 8.47\\
& Hail & 4 & 2 & 45.28 & 8.70\\
\colrule
Spring & Baseline & --- & --- & 71.83 & 4.91\\
& Heavy Rain & 53 & 23 & 64.08 & 4.86\\
& Snowstorm & 62 & 13 & 54.05 & 11.25\\
& Freezing Rain & 9 & 3 & 45.87 & 18.40\\
& Hail & 13 & 4 & 50.72 & 21.54\\
\botrule
\end{tabular*}
\label{tab:weather_impact_detailed}
\end{table}

\clearpage



\bibliography{refs}

\providecommand{\noopsort}[1]{}
\begin{thebibliography}{44}
\expandafter\ifx\csname natexlab\endcsname\relax\def\natexlab#1{#1}\fi
\providecommand{\url}[1]{\texttt{#1}}
\providecommand{\href}[2]{#2}
\providecommand{\path}[1]{#1}
\providecommand{\DOIprefix}{doi:}
\providecommand{\ArXivprefix}{arXiv:}
\providecommand{\URLprefix}{URL: }
\providecommand{\Pubmedprefix}{pmid:}
\providecommand{\doi}[1]{\href{http://dx.doi.org/#1}{\path{#1}}}
\providecommand{\Pubmed}[1]{\href{pmid:#1}{\path{#1}}}
\providecommand{\bibinfo}[2]{#2}
\ifx\xfnm\relax \def\xfnm[#1]{\unskip,\space#1}\fi
\bibitem[{Ball et~al.(2010)Ball, Barnhart, Dresner, Hansen, Neels, Odoni,
  Peterson, Sherry, Trani and Zou}]{Balletal10}
\bibinfo{author}{Ball, M.}, \bibinfo{author}{Barnhart, C.},
  \bibinfo{author}{Dresner, M.}, \bibinfo{author}{Hansen, M.},
  \bibinfo{author}{Neels, K.}, \bibinfo{author}{Odoni, A.},
  \bibinfo{author}{Peterson, E.}, \bibinfo{author}{Sherry, L.},
  \bibinfo{author}{Trani, A.}, \bibinfo{author}{Zou, B.}, \bibinfo{year}{2010}.
\newblock \bibinfo{title}{Total delay impact study: a comprehensive assessment
  of the costs and impacts of flight delay in the {U}nited {S}tates}.
\newblock \bibinfo{type}{Technical Report}. National Center for Excellence for
  Aviation Operations Research (U.S.). \bibinfo{address}{Institute of
  Transportation Studies, University of California, Berkeley, CA, USA}.
\newblock \URLprefix \url{https://rosap.ntl.bts.gov/view/dot/6234}.
\bibitem[{Beatty et~al.(1999)Beatty, Hsu, Berry and Rome}]{BeatHsuBerrRome99}
\bibinfo{author}{Beatty, R.}, \bibinfo{author}{Hsu, R.},
  \bibinfo{author}{Berry, L.}, \bibinfo{author}{Rome, J.},
  \bibinfo{year}{1999}.
\newblock \bibinfo{title}{Preliminary evaluation of flight delay propagation
  through an airline schedule}.
\newblock \bibinfo{journal}{Air Traffic Control Quarterly} \bibinfo{volume}{7},
  \bibinfo{pages}{259--270}.
\newblock \DOIprefix\doi{10.2514/atcq.7.4.259}.
\bibitem[{Beltman et~al.(2025)Beltman, Ribeiro, {de Wilde} and
  Sun}]{BeltRibedeWiSun25}
\bibinfo{author}{Beltman, M.}, \bibinfo{author}{Ribeiro, M.},
  \bibinfo{author}{{de Wilde}, J.}, \bibinfo{author}{Sun, J.},
  \bibinfo{year}{2025}.
\newblock \bibinfo{title}{Dynamically forecasting airline departure delay
  probability distributions for individual flights using supervised learning}.
\newblock \bibinfo{journal}{Journal of Air Transport Management}
  \bibinfo{volume}{126}, \bibinfo{pages}{102788}.
\newblock \DOIprefix\doi{10.1016/j.jairtraman.2025.102788}.
\bibitem[{Breiman(2001)}]{Brei01}
\bibinfo{author}{Breiman, L.}, \bibinfo{year}{2001}.
\newblock \bibinfo{title}{Random forests}.
\newblock \bibinfo{journal}{Machine Learning} \bibinfo{volume}{45},
  \bibinfo{pages}{5--32}.
\bibitem[{Bubalo and Gaggero(2021)}]{BubaGagg21}
\bibinfo{author}{Bubalo, B.}, \bibinfo{author}{Gaggero, A.A.},
  \bibinfo{year}{2021}.
\newblock \bibinfo{title}{Flight delays in {E}uropean airline networks}.
\newblock \bibinfo{journal}{Research in Transportation Business \& Management}
  \bibinfo{volume}{41}, \bibinfo{pages}{100631}.
\newblock \DOIprefix\doi{10.1016/j.rtbm.2021.100631}.
\bibitem[{Buxi and Hansen(2013)}]{BuxiHans13}
\bibinfo{author}{Buxi, G.}, \bibinfo{author}{Hansen, M.}, \bibinfo{year}{2013}.
\newblock \bibinfo{title}{Generating day-of-operation probabilistic capacity
  scenarios from weather forecasts}.
\newblock \bibinfo{journal}{Transportation Research Part C: Emerging
  Technologies} \bibinfo{volume}{33}, \bibinfo{pages}{153--166}.
\newblock \DOIprefix\doi{10.1016/j.trc.2012.12.006}.
\bibitem[{Calzada and Fageda(2023)}]{CalzFage23}
\bibinfo{author}{Calzada, J.}, \bibinfo{author}{Fageda, X.},
  \bibinfo{year}{2023}.
\newblock \bibinfo{title}{Airport dominance, route network design and flight
  delays}.
\newblock \bibinfo{journal}{Transportation Research Part E: Logistics and
  Transportation Review} \bibinfo{volume}{170}, \bibinfo{pages}{103000}.
\newblock \DOIprefix\doi{10.1016/j.tre.2022.103000}.
\bibitem[{Carvalho et~al.(2021)Carvalho, Sternberg, Gonçalves, Cruz, Soares,
  Brandão, Carvalho and Ogasawara}]{Carvetal21}
\bibinfo{author}{Carvalho, L.}, \bibinfo{author}{Sternberg, A.},
  \bibinfo{author}{Gonçalves, L.M.}, \bibinfo{author}{Cruz, A.B.},
  \bibinfo{author}{Soares, J.A.}, \bibinfo{author}{Brandão, D.},
  \bibinfo{author}{Carvalho, D.}, \bibinfo{author}{Ogasawara, E.},
  \bibinfo{year}{2021}.
\newblock \bibinfo{title}{On the relevance of data science for flight delay
  research: a systematic review}.
\newblock \bibinfo{journal}{Transport Reviews}
  \DOIprefix\doi{10.1080/01441647.2020.1861123}.
\bibitem[{Chandramouleeswaran and Tran(2018)}]{ChanTran18}
\bibinfo{author}{Chandramouleeswaran, K.R.}, \bibinfo{author}{Tran, H.T.},
  \bibinfo{year}{2018}.
\newblock \bibinfo{title}{Data-driven resilience quantification of the us air
  transportation network}, in: \bibinfo{booktitle}{2018 Annual IEEE
  International Systems Conference (SysCon)}, pp. \bibinfo{pages}{1--7}.
\newblock \DOIprefix\doi{10.1109/SYSCON.2018.8369602}.
\bibitem[{Chen and Guestrin(2016)}]{ChenGues16}
\bibinfo{author}{Chen, T.}, \bibinfo{author}{Guestrin, C.},
  \bibinfo{year}{2016}.
\newblock \bibinfo{title}{{XGBoost}: A scalable tree boosting system}, in:
  \bibinfo{booktitle}{Proceedings of the 22nd ACM SIGKDD International
  Conference on Knowledge Discovery and Data Mining},
  \bibinfo{publisher}{Association for Computing Machinery},
  \bibinfo{address}{New York, NY, USA}. pp. \bibinfo{pages}{785--794}.
\newblock \DOIprefix\doi{10.1145/2939672.2939785}.
\bibitem[{Chin et~al.(2013)Chin, Meilus, Murphy and
  Thyagarajan}]{ChinMeilMurpThya13}
\bibinfo{author}{Chin, D.K.}, \bibinfo{author}{Meilus, A.J.},
  \bibinfo{author}{Murphy, D.}, \bibinfo{author}{Thyagarajan, P.},
  \bibinfo{year}{2013}.
\newblock \bibinfo{title}{Forecasting airport delays}, in:
  \bibinfo{editor}{Andreatta, G.}, \bibinfo{editor}{Odoni, A.R.} (Eds.),
  \bibinfo{booktitle}{Modelling and Managing Airport Performance}.
  \bibinfo{publisher}{Wiley}, \bibinfo{address}{Chichester, West Sussex, United
  Kingdom}.
\newblock \DOIprefix\doi{10.1002/9781118535844.ch4}.
\bibitem[{Dai(2024)}]{Dai24}
\bibinfo{author}{Dai, M.}, \bibinfo{year}{2024}.
\newblock \bibinfo{title}{A hybrid machine learning-based model for predicting
  flight delay through aviation big data}.
\newblock \bibinfo{journal}{Scientific Reports} \bibinfo{volume}{14}.
\newblock \DOIprefix\doi{10.1038/s41598-024-55217-z}.
\bibitem[{Etani(2019)}]{Etan19}
\bibinfo{author}{Etani, N.}, \bibinfo{year}{2019}.
\newblock \bibinfo{title}{Development of a predictive model for on-time arrival
  flight of airliner by discovering correlation between flight and weather
  data}.
\newblock \bibinfo{journal}{Journal of Big Data} \bibinfo{volume}{6}.
\newblock \DOIprefix\doi{10.1186/s40537-019-0251-y}.
\bibitem[{Fisher et~al.(2019)Fisher, Rudin and Dominici}]{FishRudiDomi19}
\bibinfo{author}{Fisher, A.}, \bibinfo{author}{Rudin, C.},
  \bibinfo{author}{Dominici, F.}, \bibinfo{year}{2019}.
\newblock \bibinfo{title}{All models are wrong, but many are useful: Learning a
  variable’s importance by studying an entire class of prediction models
  simultaneously}.
\newblock \bibinfo{journal}{Journal of Machine Learning Research}
  \bibinfo{volume}{20}.
\bibitem[{Friedman(2001)}]{Frie01}
\bibinfo{author}{Friedman, J.H.}, \bibinfo{year}{2001}.
\newblock \bibinfo{title}{Greedy function approximation: A gradient boosting
  machine}.
\newblock \bibinfo{journal}{Annals of Statistics} \bibinfo{volume}{29},
  \bibinfo{pages}{1189--1232}.
\newblock \DOIprefix\doi{10.1214/aos/1013203451}.
\bibitem[{Friedman(2002)}]{Frie02}
\bibinfo{author}{Friedman, J.H.}, \bibinfo{year}{2002}.
\newblock \bibinfo{title}{Stochastic gradient boosting}.
\newblock \bibinfo{journal}{Computational Statistics \& Data Analysis}
  \bibinfo{volume}{38}, \bibinfo{pages}{367--378}.
\newblock \DOIprefix\doi{10.1016/S0167-9473(01)00065-2}.
  \bibinfo{note}{nonlinear Methods and Data Mining}.
\bibitem[{Guo et~al.(2021)Guo, Yu, Hao, Wang, Jiang and
  Zong}]{GuoYuHaoWangJianZong21}
\bibinfo{author}{Guo, Z.}, \bibinfo{author}{Yu, B.}, \bibinfo{author}{Hao, M.},
  \bibinfo{author}{Wang, W.}, \bibinfo{author}{Jiang, Y.},
  \bibinfo{author}{Zong, F.}, \bibinfo{year}{2021}.
\newblock \bibinfo{title}{A novel hybrid method for flight departure delay
  prediction using random forest regression and maximal information
  coefficient}.
\newblock \bibinfo{journal}{Aerospace Science and Technology}
  \bibinfo{volume}{116}, \bibinfo{pages}{106822}.
\newblock \DOIprefix\doi{10.1016/j.ast.2021.106822}.
\bibitem[{Hajko and Badánik(2020)}]{HajkBada20}
\bibinfo{author}{Hajko, J.}, \bibinfo{author}{Badánik, B.},
  \bibinfo{year}{2020}.
\newblock \bibinfo{title}{Airline on-time performance management}.
\newblock \bibinfo{journal}{Transportation Research Procedia}
  \bibinfo{volume}{51}, \bibinfo{pages}{82--97}.
\newblock \DOIprefix\doi{10.1016/j.trpro.2020.11.011}. \bibinfo{note}{iNAIR
  2020 - challenges of aviation development}.
\bibitem[{Khan et~al.(2021)Khan, Ma, Chung and Wen}]{KhanMaChunWen21}
\bibinfo{author}{Khan, W.A.}, \bibinfo{author}{Ma, H.L.},
  \bibinfo{author}{Chung, S.H.}, \bibinfo{author}{Wen, X.},
  \bibinfo{year}{2021}.
\newblock \bibinfo{title}{Hierarchical integrated machine learning model for
  predicting flight departure delays and duration in series}.
\newblock \bibinfo{journal}{Transportation Research Part C: Emerging
  Technologies} \bibinfo{volume}{129}, \bibinfo{pages}{103225}.
\newblock \DOIprefix\doi{10.1016/j.trc.2021.103225}.
\bibitem[{Kim and Park(2016)}]{KimPark16}
\bibinfo{author}{Kim, N.Y.}, \bibinfo{author}{Park, J.W.},
  \bibinfo{year}{2016}.
\newblock \bibinfo{title}{A study on the impact of airline service delays on
  emotional reactions and customer behavior}.
\newblock \bibinfo{journal}{Journal of Air Transport Management}
  \bibinfo{volume}{57}, \bibinfo{pages}{19--25}.
\newblock \DOIprefix\doi{10.1016/j.jairtraman.2016.07.005}.
\bibitem[{Li and Yao(2025)}]{LiYao25}
\bibinfo{author}{Li, N.}, \bibinfo{author}{Yao, H.G.}, \bibinfo{year}{2025}.
\newblock \bibinfo{title}{A review of research on flight delay propagation:
  Current situation and prospect}.
\newblock \bibinfo{journal}{Journal of Advanced Transportation}
  \bibinfo{volume}{2025}, \bibinfo{pages}{4851103}.
\newblock \DOIprefix\doi{10.1155/atr/4851103}.
\bibitem[{Li and Jing(2022)}]{LiJing22}
\bibinfo{author}{Li, Q.}, \bibinfo{author}{Jing, R.}, \bibinfo{year}{2022}.
\newblock \bibinfo{title}{Flight delay prediction from spatial and temporal
  perspective}.
\newblock \bibinfo{journal}{Expert Systems with Applications}
  \bibinfo{volume}{205}, \bibinfo{pages}{117662}.
\newblock \DOIprefix\doi{10.1016/j.eswa.2022.117662}.
\bibitem[{Liaw and Wiener(2002)}]{LiawWien02}
\bibinfo{author}{Liaw, A.}, \bibinfo{author}{Wiener, M.}, \bibinfo{year}{2002}.
\newblock \bibinfo{title}{Classification and regression by r{andomForest}}.
\newblock \bibinfo{journal}{R News} \bibinfo{volume}{2},
  \bibinfo{pages}{18--22}.
\bibitem[{Lundberg and Lee(2017)}]{LundLee17}
\bibinfo{author}{Lundberg, S.M.}, \bibinfo{author}{Lee, S.I.},
  \bibinfo{year}{2017}.
\newblock \bibinfo{title}{A unified approach to interpreting model
  predictions}, in: \bibinfo{booktitle}{Proceedings of the 31st International
  Conference on Neural Information Processing Systems},
  \bibinfo{publisher}{Curran Associates Inc.}, \bibinfo{address}{Red Hook, NY,
  USA}. pp. \bibinfo{pages}{4768--4777}.
\bibitem[{Manchiraju et~al.(2023)Manchiraju, Sohoni and
  Deshpande}]{MancSohoDesh23}
\bibinfo{author}{Manchiraju, C.}, \bibinfo{author}{Sohoni, M.G.},
  \bibinfo{author}{Deshpande, V.}, \bibinfo{year}{2023}.
\newblock \bibinfo{title}{It's not simply luck: The impact of network strategy,
  schedule padding, and operational improvements on domestic on-time
  performance in the us airline industry}.
\newblock \bibinfo{journal}{Production and Operations Management}
  \bibinfo{volume}{32}, \bibinfo{pages}{3559--3574}.
\newblock \DOIprefix\doi{10.1111/poms.14050}.
\bibitem[{{OAG}(2024)}]{OAG24}
\bibinfo{author}{{OAG}}, \bibinfo{year}{2024}.
\newblock \bibinfo{title}{Megahubs 2024}.
\newblock \URLprefix
  \url{https://www.oag.com/hubfs/Megahubs%202024/Megahubs-2024.pdf}.
  \bibinfo{note}{accessed: 2025-09-30}.
\bibitem[{Pedregosa et~al.(2011)Pedregosa, Varoquaux, Gramfort, Michel,
  Thirion, Grisel, Blondel, Prettenhofer, Weiss, Dubourg, Vanderplas, Passos,
  Cournapeau, Brucher, Perrot and Duchesnay}]{Pedretal11}
\bibinfo{author}{Pedregosa, F.}, \bibinfo{author}{Varoquaux, G.},
  \bibinfo{author}{Gramfort, A.}, \bibinfo{author}{Michel, V.},
  \bibinfo{author}{Thirion, B.}, \bibinfo{author}{Grisel, O.},
  \bibinfo{author}{Blondel, M.}, \bibinfo{author}{Prettenhofer, P.},
  \bibinfo{author}{Weiss, R.}, \bibinfo{author}{Dubourg, V.},
  \bibinfo{author}{Vanderplas, J.}, \bibinfo{author}{Passos, A.},
  \bibinfo{author}{Cournapeau, D.}, \bibinfo{author}{Brucher, M.},
  \bibinfo{author}{Perrot, M.}, \bibinfo{author}{Duchesnay, {\'E}.},
  \bibinfo{year}{2011}.
\newblock \bibinfo{title}{Scikit-learn: Machine learning in {P}ython}.
\newblock \bibinfo{journal}{Journal of Machine Learning Research}
  \bibinfo{volume}{12}, \bibinfo{pages}{2825--2830}.
\bibitem[{{Performance Review Commission}(2005)}]{Euro05}
\bibinfo{author}{{Performance Review Commission}}, \bibinfo{year}{2005}.
\newblock \bibinfo{title}{Report on Punctuality Drivers at Major European
  Airports}.
\newblock \bibinfo{type}{Technical Report}. EUROCONTROL.
  \bibinfo{address}{Brussels, Belgium}.
\bibitem[{Pyrgiotis et~al.(2013)Pyrgiotis, Malone and Odoni}]{PyrgMaloOdon13}
\bibinfo{author}{Pyrgiotis, N.}, \bibinfo{author}{Malone, K.M.},
  \bibinfo{author}{Odoni, A.}, \bibinfo{year}{2013}.
\newblock \bibinfo{title}{Modelling delay propagation within an airport
  network}.
\newblock \bibinfo{journal}{Transportation Research Part C}
  \bibinfo{volume}{27}, \bibinfo{pages}{60--75}.
\newblock \DOIprefix\doi{10.1016/j.trc.2011.05.017}.
\bibitem[{Rebollo and Balakrishnan(2014)}]{ReboBala14}
\bibinfo{author}{Rebollo, J.J.}, \bibinfo{author}{Balakrishnan, H.},
  \bibinfo{year}{2014}.
\newblock \bibinfo{title}{Characterization and prediction of air traffic
  delays}.
\newblock \bibinfo{journal}{Transportation Research Part C: Emerging
  Technologies} \bibinfo{volume}{44}, \bibinfo{pages}{231--241}.
\newblock \DOIprefix\doi{10.1016/j.trc.2014.04.007}.
\bibitem[{Ribeiro et~al.(2025)Ribeiro, Tay, Ng and Birolini}]{RibeTayNgBiro25}
\bibinfo{author}{Ribeiro, N.A.}, \bibinfo{author}{Tay, J.},
  \bibinfo{author}{Ng, W.}, \bibinfo{author}{Birolini, S.},
  \bibinfo{year}{2025}.
\newblock \bibinfo{title}{Delay predictive analytics for airport capacity
  management}.
\newblock \bibinfo{journal}{Transportation Research Part C: Emerging
  Technologies} \bibinfo{volume}{171}, \bibinfo{pages}{104947}.
\newblock \DOIprefix\doi{10.1016/j.trc.2024.104947}.
\bibitem[{Álvaro Rodríguez-Sanz et~al.(2019)Álvaro Rodríguez-Sanz,
  Gómez~Comendador, Arnaldo~Valdés, Pérez-Castán, Barragán~Montes and
  Cámara~Serrano}]{RodrGomeArnaPereBarr19}
\bibinfo{author}{Álvaro Rodríguez-Sanz}, \bibinfo{author}{Gómez~Comendador,
  F.}, \bibinfo{author}{Arnaldo~Valdés, R.}, \bibinfo{author}{Pérez-Castán,
  J.}, \bibinfo{author}{Barragán~Montes, R.},
  \bibinfo{author}{Cámara~Serrano, S.}, \bibinfo{year}{2019}.
\newblock \bibinfo{title}{Assessment of airport arrival congestion and delay:
  Prediction and reliability}.
\newblock \bibinfo{journal}{Transportation Research Part C}
  \bibinfo{volume}{98}, \bibinfo{pages}{255--283}.
\newblock \DOIprefix\doi{10.1016/j.trc.2018.11.015}.
\bibitem[{Rudin et~al.(2022)Rudin, Chen, Chen, Huang, Semenova and
  Zhong}]{RudiChenChenHuanSemeZhon22}
\bibinfo{author}{Rudin, C.}, \bibinfo{author}{Chen, C.}, \bibinfo{author}{Chen,
  Z.}, \bibinfo{author}{Huang, H.}, \bibinfo{author}{Semenova, L.},
  \bibinfo{author}{Zhong, C.}, \bibinfo{year}{2022}.
\newblock \bibinfo{title}{Interpretable machine learning: Fundamental
  principles and 10 grand challenges}.
\newblock \bibinfo{journal}{Statistics Surveys} \bibinfo{volume}{16},
  \bibinfo{pages}{1--85}.
\newblock \DOIprefix\doi{10.1214/21-SS133}.
\bibitem[{Strobl et~al.(2007)Strobl, Boulesteix, Zeileis and
  Hothorn}]{StroBoulZeilHoth07}
\bibinfo{author}{Strobl, C.}, \bibinfo{author}{Boulesteix, A.L.},
  \bibinfo{author}{Zeileis, A.}, \bibinfo{author}{Hothorn, T.},
  \bibinfo{year}{2007}.
\newblock \bibinfo{title}{Bias in random forest variable importance measures:
  Illustrations, sources and a solution}.
\newblock \bibinfo{journal}{BMC Bioinformatics} \bibinfo{volume}{8}.
\newblock \DOIprefix\doi{10.1186/1471-2105-8-25}.
\bibitem[{Tan et~al.(2025)Tan, Lu and Wang}]{TanLuWang25}
\bibinfo{author}{Tan, Y.}, \bibinfo{author}{Lu, Y.}, \bibinfo{author}{Wang,
  L.}, \bibinfo{year}{2025}.
\newblock \bibinfo{title}{Flight delay dynamics: Unraveling the impact of
  airport-network-spilled propagation on airline on-time performance}.
\newblock \bibinfo{journal}{Decision Support Systems} \bibinfo{volume}{196},
  \bibinfo{pages}{114494}.
\newblock \DOIprefix\doi{10.1016/j.dss.2025.114494}.
\bibitem[{Wandelt et~al.(2025)Wandelt, Chen and Sun}]{WandChenSun25}
\bibinfo{author}{Wandelt, S.}, \bibinfo{author}{Chen, X.},
  \bibinfo{author}{Sun, X.}, \bibinfo{year}{2025}.
\newblock \bibinfo{title}{Flight delay prediction: A dissecting review of
  recent studies using machine learning}.
\newblock \bibinfo{journal}{IEEE Transactions on Intelligent Transportation
  Systems} \bibinfo{volume}{26}, \bibinfo{pages}{4283--4297}.
\newblock \DOIprefix\doi{10.1109/TITS.2025.3528536}.
\bibitem[{Wang et~al.(2022a)Wang, Bi, Xie and Zhao}]{WangBiXieZhao22}
\bibinfo{author}{Wang, F.}, \bibinfo{author}{Bi, J.}, \bibinfo{author}{Xie,
  D.}, \bibinfo{author}{Zhao, X.}, \bibinfo{year}{2022}a.
\newblock \bibinfo{title}{Flight delay forecasting and analysis of direct and
  indirect factors}.
\newblock \bibinfo{journal}{IET Intelligent Transport Systems}
  \bibinfo{volume}{16}, \bibinfo{pages}{890--907}.
\newblock \DOIprefix\doi{10.1049/itr2.12183}.
\bibitem[{Wang et~al.(2022b)Wang, Allignol, Barnier, Gondran, Gotteland and
  Mancel}]{WangAlliBarnGondGottManc22}
\bibinfo{author}{Wang, R.}, \bibinfo{author}{Allignol, C.},
  \bibinfo{author}{Barnier, N.}, \bibinfo{author}{Gondran, A.},
  \bibinfo{author}{Gotteland, J.B.}, \bibinfo{author}{Mancel, C.},
  \bibinfo{year}{2022}b.
\newblock \bibinfo{title}{A new multi-commodity flow model to optimize the
  robustness of the gate allocation problem}.
\newblock \bibinfo{journal}{Transportation Research Part C: Emerging
  Technologies} \bibinfo{volume}{136}, \bibinfo{pages}{103491}.
\newblock \DOIprefix\doi{10.1016/j.trc.2021.103491}.
\bibitem[{Xu et~al.(2008)Xu, Sherry and Laskey}]{XuSherLask08}
\bibinfo{author}{Xu, N.}, \bibinfo{author}{Sherry, L.},
  \bibinfo{author}{Laskey, K.B.}, \bibinfo{year}{2008}.
\newblock \bibinfo{title}{Multifactor model for predicting delays at u.s.
  airports}.
\newblock \bibinfo{journal}{Transportation Research Record}
  \bibinfo{volume}{2052}, \bibinfo{pages}{62--71}.
\newblock \DOIprefix\doi{10.3141/2052-08},
  \href{http://arxiv.org/abs/https://doi.org/10.3141/2052-08}{{\tt
  arXiv:https://doi.org/10.3141/2052-08}}.
\bibitem[{Yang et~al.(2023)Yang, Chen, Hu, Song and Mao}]{YangChenHuSongMao23}
\bibinfo{author}{Yang, Z.}, \bibinfo{author}{Chen, Y.}, \bibinfo{author}{Hu,
  J.}, \bibinfo{author}{Song, Y.}, \bibinfo{author}{Mao, Y.},
  \bibinfo{year}{2023}.
\newblock \bibinfo{title}{Departure delay prediction and analysis based on node
  sequence data of ground support services for transit flights}.
\newblock \bibinfo{journal}{Transportation Research Part C}
  \bibinfo{volume}{153}, \bibinfo{pages}{104217}.
\newblock \DOIprefix\doi{10.1016/j.trc.2023.104217}.
\bibitem[{Yimga(2017)}]{Yimg17}
\bibinfo{author}{Yimga, J.}, \bibinfo{year}{2017}.
\newblock \bibinfo{title}{Airline on-time performance and its effects on
  consumer choice behavior}.
\newblock \bibinfo{journal}{Research in Transportation Economics}
  \bibinfo{volume}{66}, \bibinfo{pages}{12--25}.
\newblock \DOIprefix\doi{10.1016/j.retrec.2017.06.001}.
\bibitem[{Yimga and Gorjidooz(2019)}]{YimgGorj19}
\bibinfo{author}{Yimga, J.}, \bibinfo{author}{Gorjidooz, J.},
  \bibinfo{year}{2019}.
\newblock \bibinfo{title}{Airline schedule padding and consumer choice
  behavior}.
\newblock \bibinfo{journal}{Journal of Air Transport Management}
  \bibinfo{volume}{78}, \bibinfo{pages}{71--79}.
\newblock \DOIprefix\doi{10.1016/j.jairtraman.2019.05.001}.
\bibitem[{Zheng et~al.(2024)Zheng, Wang, Zheng, Wang, Fan, Cong and
  Hu}]{Zhenetal24}
\bibinfo{author}{Zheng, H.}, \bibinfo{author}{Wang, Z.},
  \bibinfo{author}{Zheng, C.}, \bibinfo{author}{Wang, Y.},
  \bibinfo{author}{Fan, X.}, \bibinfo{author}{Cong, W.}, \bibinfo{author}{Hu,
  M.}, \bibinfo{year}{2024}.
\newblock \bibinfo{title}{A graph multi-attention network for predicting
  airport delays}.
\newblock \bibinfo{journal}{Transportation Research Part E: Logistics and
  Transportation Review} \bibinfo{volume}{181}, \bibinfo{pages}{103375}.
\newblock \DOIprefix\doi{10.1016/j.tre.2023.103375}.
\bibitem[{Zhong et~al.(2025)Zhong, Yu, Huang and Zhang}]{ZhonYuHuanZhan25}
\bibinfo{author}{Zhong, Q.}, \bibinfo{author}{Yu, Y.}, \bibinfo{author}{Huang,
  Y.}, \bibinfo{author}{Zhang, T.}, \bibinfo{year}{2025}.
\newblock \bibinfo{title}{Prediction and optimization of civil aviation flight
  delays based on machine learning algorithms}.
\newblock \bibinfo{journal}{International Journal of Computational Intelligence
  Systems} \bibinfo{volume}{18}.

\end{thebibliography}
\bibliographystyle{elsarticle-harv}



\end{document}